\documentclass[aip,cha,reprint]{revtex4-1}
\usepackage{mathtools}
\usepackage{xcolor}
\usepackage{graphicx}
\usepackage{amsmath}
\usepackage{amssymb}
\usepackage[hidelinks]{hyperref}
\usepackage{fancyhdr}

\fancypagestyle{firstpage}{ % first page style
	
	\fancyhead{\fontsize{10}{15}\selectfont Chaos {\bf 32}, 113126 (2022);\quad doi: 10.1063/5.0106171}
	\fancyhead[L]{}
	\fancyhead[R]{}
	\fancyfoot[C]{}
}

\DeclareMathOperator\arctanh{arctanh}

\def\e{\varepsilon}
\def\vp{\varphi}
\newcommand{\av}[1]{\left\langle #1 \right\rangle}
\def\ii{\mathrm{i}}

\begin{document}

\title{Exact finite-dimensional reduction for a population of noisy oscillators\\ and its link to Ott-Antonsen and Watanabe-Strogatz theories}

\author{Rok Cestnik}
%\email{rokcestn@uni-potsdam.de}
\affiliation{Department of Physics and Astronomy, University of Potsdam, Karl-Liebknecht-Strasse 24/25, 14476, Potsdam-Golm, Germany}
\author{Arkady Pikovsky}
\affiliation{Department of Physics and Astronomy, University of Potsdam, Karl-Liebknecht-Strasse 24/25, 14476, Potsdam-Golm, Germany}

\begin{abstract}
Populations of globally coupled phase oscillators are described in the thermodynamic limit by kinetic equations
for the distribution densities, or equivalently, by infinite hierarchies of equations for the order parameters.
Ott and Antonsen [Chaos 18, 037113 (2008)] have found an invariant finite-dimensional subspace on which the dynamics
is described by one complex variable per population. For oscillators with Cauchy distributed frequencies or for those
driven by Cauchy white noise, this subspace is weakly stable and thus describes the asymptotic dynamics.
Here we report on an exact finite-dimensional reduction of the dynamics outside of the Ott-Antonsen subspace.
We show, that the evolution from generic initial states can be reduced to that of three complex variables, plus
a constant function. For identical noise-free oscillators, this reduction corresponds to the Watanabe-Strogatz
system of equations [Phys. Rev. Lett. 70, 2391 (1993)]. We discuss how the reduced system can be used to explore the transient dynamics
of perturbed ensembles.
\end{abstract}

\maketitle
\thispagestyle{firstpage}

\begin{quotation}
Large ensembles of globally coupled oscillators can be described by means of kinetic equations
for the evolution of the distribution densities. These equations take the simplest form
if the oscillators are described by their phases only. Alternatively, one can write an infinite
set of ordinary differential equations for the set of order parameters (Fourier modes of the distribution).
For oscillators driven by Cauchy white noise or for a Cauchy distribution of natural frequencies, the system
of equations for the order parameters (for coupling in the first harmonics of the phase),
has a remarkable property first discovered by Ott and Antonsen in 2008: it possesses an invariant manifold
on which the dynamics reduces to just one complex equation. In this paper we extend this result by showing, that
for arbitrary initial conditions the dynamics reduces to that of three complex variables. 
In the noise-free case of identical oscillators, our equations are equivalent to the system derived 
by Watanabe and Strogatz in 1993. The finite-dimensional reduction allows 
exact calculation of transients 
to the attracting Ott-Antonsen states,
by solving a simple six-dimensional system of equations.   
\end{quotation}

\section{Introduction}
Ensembles of coupled oscillators is a popular object in studies of complex systems,
with a wide range of applications; from physical systems (lasers~\cite{Nixon_etal-13}, 
Josephson junctions~\cite{Wiesenfeld-Swift-95,Wiesenfeld-Colet-Strogatz-98}, chemical reactions~\cite{Totz_etal-18}), to engineering (pedestrians on a bridge~\cite{Eckhardt_et_al-07}) and life sciences (neurons~\cite{Luke-Barreto-So-13,Laing-14}, nephron cells~\cite{Holstein_etal-01}, genetic circuits~\cite{Prindle_etal-12}). A common  theoretical approach
includes different levels of reductions and idealizations. If the units are self-sustained periodic oscillators, and the coupling is weak, one can perform a phase reduction, neglecting variations of the oscillators' amplitudes that appear in the higher orders in coupling strength~\cite{Kuramoto-84}. As a result, each oscillator is described by just one variable on a unit circle -- the phase, which enormously simplifies the analysis. Another idealization, which is appropriate for large ensembles, is the thermodynamic limit of an infinite
number of units. This allows for a formulation of the evolution
in terms of kinetic equations for the distribution of the phases. An important class of models are those with 
global (or mean-field) coupling. Such models appear naturally, e.g.,  for Josephson junctions with a common load and for pedestrians on a bridge; in other cases (e.g., for neural ensembles) they
are justified by a huge number of interconnections between the units.

Among the setups for ensembles of globally coupled phase oscillators, the paradigmatic Kuramoto
model~\cite{kuramoto_model} and its generalizations~\cite{Sakaguchi-Kuramoto-86,Acebron-etal-05} are particularly popular. Here one assumes a relatively
simple coupling, where the dynamics of the oscillator's phase depends only on the first   harmonics of
the phase itself. To define the coupling, one introduces mean fields which are the circular moments
of the phase distribution. Different setups with identical deterministic units, as well as ones having different natural frequencies and/or being driven by noise have been considered in the literature.

One of the striking properties of the Kuramoto-type models is the possibility to reduce the dynamics
to a finite-dimensional one. Watanabe and 
Strogatz~\cite{watanabe_strogatz_1993,Watanabe-Strogatz-94} (WS) have demonstrated that ensembles
of identical, noise-free units can be exactly reduced to three dynamical equations (plus constants of motion). Ott and Antonsen~\cite{ott_antonsen_2008} (OA) found a particular family of phase distributions (wrapped Cauchy distribution) that is invariant under the dynamical evolution. This holds not only for identical units, but also
for ones with a Cauchy distribution of natural frequencies, 
and for ones driven by white Cauchy
noise~\cite{Tanaka-20,tonjes_pikovsky_2020}. In contradistinction to WS theory, the OA reduction is not valid for arbitrary initial states - they should belong to the OA  invariant manifold. However, because there are arguments that the OA manifold is attracting (although not in a trivial sense, see 
discussion in \cite{Ott-Antonsen-09,pietras_daffertshofer_2016,engelbrecht_mirollo_2020}), the OA equations correctly
describe the asymptotic in time regimes.

The goal of this paper is to fill, at least partially, the gap between WS and OA theories. We will develop,
in the thermodynamic limit,
a low-dimensional description of the Kuramoto-type phase ensembles with Cauchy noise and/or Cauchy distribution
of natural frequencies, valid for arbitrary initial conditions. Of course, this reduction contains WS and OA equations as particular cases. 

The paper is organized as follows. In section \ref{sec:pf} we formulate the problem. 
In section \ref{sec:fdr} we introduce our basic tools (generating functions) and 
define a family of finite-dimensional invariant manifolds (these results have been also
presented in a short communication~\cite{cestnik_pikovsky_2022}). Section \ref{sec:tcv} contains the main result - we show how the evolution of generic states can be reduced to three complex variables plus a constant function.
Here we also discuss different possibilities of introducing these variables based on initial conditions.
In section \ref{sec:gsoa} we demonstrate stability of the OA manifold in the presence of noise. 
In section \ref{sec:nfc} we consider identical noise-free oscillators, and demonstrate that
the dynamics reduces to the WS equations. Section \ref{sec:ls} is devoted to the implications for the spectrum of the Lyapunov exponents.
In section \ref{sec:res} we discuss how our approach allows for finding the 
evolution outside of the OA manifold.   
We conclude and discuss possible further developments in section~\ref{sec:concl}. 
Many technical details are shifted from the main text to appendices.

\section{Problem formulation}
\label{sec:pf}
In this paper we consider populations of phase oscillators, subject to global coupling or to
a global common force, in the thermodynamic limit
of an infinite number of units. Consequently, the proper description is in terms of the phase distribution functions.
Our theory is valid for a restricted class of systems: (i) important is that the coupling/forcing is proportional 
to the first harmonics of the phase only, and (ii) the oscillators can differ from each other only in additive terms
in their phase dynamics, which are either Cauchy-distributed white noise terms, or Cauchy-distributed frequency constants, or a combination of both.
In this section we introduce these models.
\subsection{Ensemble of phase oscillators with independent Cauchy noise forces}
\label{sec:pfcn}

We consider an ensemble of noisy phase oscillators coupled in the first harmonic:
\begin{equation}
\dot{\varphi}_j = \omega(t) + \text{Im}\big[2 h(t) e^{-\ii \varphi_j}\big] + \gamma \xi_j(t)\;.
\label{eq:phase_system_noise}
\end{equation}
Here $\omega$ is a combination of a natural frequency and a 
real-valued additive force,  and $h(t)$ is a complex-valued force
that includes the first harmonic of the phase. Both these quantities can potentially depend 
on the mean fields of the population, thus readily describing global coupling. There is no restriction on these forces,
e.g., they can include noise which is then the common noise for all elements of the 
population (cf. Refs.~\onlinecite{Braun-etal-12,Gong_etal-19}).
The terms $\xi_j(t)$ represent independent, normalized Cauchy white noise forces, with $\gamma$ being the real and positive 
noise strength~\cite{Chechkin_etal-03,Toenjes_etal-13,tonjes_pikovsky_2020,Tanaka-20}.  
 
We consider the thermodynamic limit of infinitely many oscillators. In this case it is natural to describe the state with the 
phase density function $P(\varphi,t)$, and express the original dynamics in terms of the continuity equation, 
a partial differential equation (PDE) where the Cauchy noise begets a term with a fractional derivative on the right-hand side:
\begin{equation}
\frac{\partial}{\partial t}P + \frac{\partial}{\partial \vp}\big( \dot{\vp}P \big) = 
\gamma \bigg|\frac{\partial}{\partial \vp} \bigg| P\;.
\label{eq:continuity}
\end{equation}
With $\left|\frac{\partial}{\partial \vp} \right|^\alpha$ one denotes an operator, which in the Fourier representation reduces
to a multiplication with $|n|^\alpha $: $\left|\frac{\partial}{\partial \vp} \right|^\alpha e^{in\vp}=-|n|^\alpha e^{in\vp}$, cf. Ref.~\onlinecite{toenjes2014spectral}. 
In this representation, the usual Gaussian noise corresponds to $\alpha=2$, while the Cauchy
noise corresponds to $\alpha=1$ (other non-integer values of $\alpha$ describe different $\alpha$-stable distributions).

The phase density is commonly expressed as a Fourier series: 
\begin{equation}
\begin{split}
P(\varphi,t) &= \frac{1}{2\pi} \Big( 1 + \sum\limits_{n=1}^\infty Z_n(t) e^{-\ii n\varphi} + \text{c.c.} \Big)\;,\\
\qquad Z_n(t) &= \langle e^{\ii n\varphi} \rangle=\int\limits_{0}^{2\pi} d\varphi \, e^{\ii n\varphi} P(\varphi,t)\;.
\end{split}
\label{eq:four}
\end{equation}
Quantities $Z_n$ represent complex order parameters, also known as Kuramoto-Daido order parameters~\cite{kuramoto_model,daido_1996}. These circular moments of the phase distribution are in fact the ``mean fields'' which may govern
the ensemble.  
In terms of these order parameters, the dynamics is represented as an infinite set of ordinary 
differential equations (ODE):
\begin{equation}
\frac{1}{n}\dot{Z}_n = (\ii\omega-\gamma) Z_n +h Z_{n-1}-h^*Z_{n+1}\;, \quad n \geq 1\;,
\label{eq:Z_dyn}
\end{equation}
(here it suffices to consider positive $n$ only, so we replace $|n|$ in the noisy term by $n$).
These equations have been discussed in Ref.~\onlinecite{tonjes_pikovsky_2020} and represent the exact 
dynamics of system~\eqref{eq:phase_system_noise} in the thermodynamic limit, without any approximation or assumption. 
Normalization of the phase density implies $Z_0\equiv1$. 

\subsection{Ensemble with a Cauchy distribution of natural frequencies}
\label{sec:pfcf}

Equivalent equations can also be derived to represent the case of Cauchy distributed natural frequencies, the situation
widely considered starting from the initial formulation by Kuramoto~\cite{Kuramoto-75,Kuramoto-84}.
In this case we consider the terms $\xi_j$ in Eq.~\eqref{eq:phase_system_noise} as constants with a 
normalized Cauchy distribution $g(\xi)=\pi^{-1}(1+\xi^2)^{-1}$. 
The total additive force $\omega(t)+\gamma\xi_j$ can be interpreted as an instantaneous frequency of oscillator $j$. 
If $\omega=\omega_0$ is a constant, then $\omega_0+\gamma \xi_j$ is the natural frequency of oscillator $j$.
Parameter $\gamma$ describes the width of the distribution of natural frequencies 
(as we will see below, this parameter
plays the same role as the strength of the Cauchy noise above, thus we use the same letter to describe it).
In our further derivation, we follow the way presented recently in Ref.~\onlinecite{engelbrecht_mirollo_2020}.
One introduces the parameter $\xi$ into the distribution of phases $P(t,\varphi;\xi)$,
and the equation for this distribution \eqref{eq:continuity} then reads
\begin{equation}
\frac{\partial}{\partial t}P\big\rvert_\xi + \frac{\partial}{\partial \vp}\left( \left[\omega+\gamma\xi-\ii he^{-\ii\vp}+\ii h^*e^{\ii \vp}\right]P\big\rvert_\xi \right) =0\;,
\label{eq:con}
\end{equation}
where we used compact notation $P\big\rvert_\xi \equiv P(\vp,t;\xi)$. 
Of interest are the order parameters (circular moments), averaged over the additions to the frequency $\xi$:
\begin{equation}
Z_n(t)=\int\limits_{-\infty}^\infty d\xi\int\limits_{0}^{2\pi} d\vp\, e^{\ii n\vp} P\big\rvert_\xi g(\xi)\;.
\label{eq:opxi}
\end{equation} 
The main assumption allowing for explicit equations for these order parameters is analyticity of the distribution
$P(\vp,t;\xi)$ in the upper halfplane of complex $\xi$. This assumption has been first introduced 
by Ott and Antonsen in their seminal paper~\cite{ott_antonsen_2008}. The main reason behind it is the possibility
to calculate the integrals via residue integration. 
Indeed, employing the residue theorem for a contour closing the upper halfplane in \eqref{eq:opxi} 
and taking the only pole at $\xi=\ii$, one reduces \eqref{eq:opxi} to $Z_n(t)=\int d\vp\, e^{\ii n\vp}P(\vp,t;\ii)$. 
Now, let us multiply \eqref{eq:con} with $e^{\ii n\vp} g(\xi)$ and integrate in  $\xi$ and $\vp$. 
The only additional integral to be calculated (again by virtue of the residue method) is 
\[
\int\limits_0^{2\pi} d\vp \int\limits_{-\infty}^\infty d\xi\; \xi e^{\ii n\vp} P\big\rvert_\xi g(\xi)=\ii \int\limits_{0}^{2\pi} d\vp \, e^{\ii n\vp} P\big\rvert_\ii=\ii Z_n\;.
\]
This yields the system of equations
\[
\dot{Z}_n=\ii n(\omega+\ii\gamma)Z_n+n h Z_{n-1}-nh^*Z_{n+1}\;,
\]
which coincides with \eqref{eq:Z_dyn}.

We end this section with two remarks. First, the validity of Eqs.~\eqref{eq:Z_dyn} for Cauchy independent noises is unconditional, while
for the Cauchy distributed constant additions to the frequency, an extra assumption of analyticity has to be adopted; the validity
of this assumption is commonly assumed in the OA theory and its applications, but one can construct  distributions of the phase which, at least
during some time interval, violate this assumption~\cite{Pikovsky-Rosenblum-11}. 

The second remark is that if one has both Cauchy-distributed constant and noisy additions to the frequency, with
intensities $\gamma_1$ and $\gamma_2$, then one can use Eqs.~\eqref{eq:Z_dyn} with the total 
intensity $\gamma=\gamma_1+\gamma_2$.

\section{Generating functions and finite-dimensional reductions of the dynamics}
\label{sec:fdr}
\subsection{Ordinary and exponential generating functions}
In our treatment of the infinite system \eqref{eq:Z_dyn} we will make use of generating functions, which are formal power series.
We will use both the ordinary generating function (OGF)~\footnote{The sum defining ordinary generating functions is typically considered from index $n=0$, but in our context it is convenient to start from $n=1$; note that we always have $f_0=1$ for normalization reasons.}
\[
\mathcal{F}(k)=\sum_{n=1}^\infty f_n k^n\;,
\]
and the exponential generating function (EGF)
\[
\mathsf{F}(k)=\sum_{n=0}^\infty f_n \frac{k^n}{n!}\;.
\]
There is no simple relation between these functions for the same sequence $\{f_n\}$, and in different situations we will use different
generating functions.

\subsection{Finite-dimensional reductions of the infinite system for circular moments}
In this section we briefly introduce the finite-dimensional reductions described in our recent letter \cite{cestnik_pikovsky_2022}.
First, we characterize the state with complex 
order parameters $Z_n$ by introducing the complex-valued EGF (because $|Z_n|\leq 1$, this series converges for all $k$): 
\begin{equation}
\mathsf{Z}(k,t) = \sum\limits_{n=0}^\infty Z_n(t) \frac{k^n}{n!}\;.
\end{equation}
Then the dynamics~\eqref{eq:Z_dyn} are recast as a single 
PDE (see Appendix \ref{ap:gf} for the derivation), which in contrast to Eq.~\eqref{eq:continuity} is generally complex (prime denotes 
derivative with respect to $k$):
\begin{equation}
\dot{\mathsf{Z}} = (\ii \omega-\gamma) k \mathsf{Z}' + h k \mathsf{Z} - h^* k \mathsf{Z}''\;.
\label{eq:F}
\end{equation}
The normalization condition is $\mathsf{Z}(0,t)=1$.
The structure of this equation allows for a particular solution with the exponential 
ansatz $\mathsf{Z}(k,t) = e^{k Q(t)}$, revealing a single ODE for the complex variable $Q(t)$:
\begin{equation}
\dot{Q} = (\ii\omega-\gamma) Q + h - h^*Q^2\;.
\label{eq:Q}
\end{equation}
This is commonly known as the Ott-Antonsen ansatz~\cite{ott_antonsen_2008}, which reveals a two-dimensional invariant manifold 
in the infinite system \eqref{eq:Z_dyn}. In this case, higher circular moments are powers of the first one: $Z_n=Q^n$. The distribution of the phases is the wrapped Cauchy
distribution (a.k.a. Poisson kernel): 
\begin{equation}
P(\vp,t) = \frac{1}{2\pi} \frac{1-|Q|^2}{|1-Qe^{-\ii \vp}|^2}\;.
\label{eq:cdist}
\end{equation} 

We recently generalized this solution with an ansatz allowing for an additional function~\cite{cestnik_pikovsky_2022}: 
\begin{equation}
\mathsf{Z}(k,t) = e^{kQ(t)}\mathsf{B}(k,t)\;, 
\label{eq:zb}
\end{equation}
in which case we obtain, in addition to \eqref{eq:Q}, another PDE for the newly 
introduced function $\mathsf{B}(k,t)$:
\begin{equation}
\dot{\mathsf{B}} = \big( \ii \omega -\gamma- 2h^*Q \big) k \mathsf{B}' - h^* k \mathsf{B}''\;.
\label{eq:bgf}
\end{equation}
Although at first glance this equation is similar to Eq.~\eqref{eq:F}, it does not contain 
a term without $k$-derivative of $\mathsf{B}(k,t)$, and thus allows for a more general dimensionality reduction. 
Namely, we expand the function $\mathsf{B}(k)$ as an 
EGF (we will see below that coefficients $\beta_n$ grow not faster than $\sim \text{const}^n$, thus this series converges for all $k$): 
\begin{equation}
\mathsf{B}(k,t) = \sum\limits_{n=0}^\infty \beta_n(t)\frac{k^n}{n!}\;,
\end{equation}
($\beta_0\equiv 1$ due to normalization), thus introducing new dynamical
variables $\beta_n(t)$, that describe the dynamics with an infinite set of ODEs 
(plus one ODE \eqref{eq:Q} for $Q$):
\begin{subequations}
\begin{align}
\dot{Q} &= (\ii\omega-\gamma) Q + h - h^* Q^2\; \label{eq:eqQ}\;,\\
\frac{1}{n}\dot{\beta}_{n} &= (\ii\omega -\gamma -2h^*Q)\beta_{n}-h^*\beta_{n+1}\;, \quad n \geq 1\; ,\label{eq:beta}
\end{align}
\label{eq:Qbeta}
\end{subequations}
(see Appendix \ref{ap:gf} for the relation of \eqref{eq:beta} to \eqref{eq:bgf}).
Notice how the right-hand side of \eqref{eq:beta} only contains terms proportional to $\beta_n$ and $\beta_{n+1}$, but
no term with $\beta_{n-1}$ is present. 
This means that if the system is truncated at a finite number $N$ of variables $\beta_n$ (i.e. assuming that all higher terms
vanish identically: $\beta_{n\geq N} = 0$), the dynamics is exactly described by the first $N$ equations of 
system~\eqref{eq:beta} for all times. These truncations represent dynamically invariant finite-dimensional manifolds.
The $\beta_n$ variables relate to the Kuramoto-Daido order parameters $Z_n$ via a modified binomial transform~(cf. Refs.~\onlinecite{number_theory_1993,number_theory_1994}):
\begin{equation}
\begin{split}
Z_n(t) &= \sum\limits_{m=0}^n \binom{n}{m} \beta_m(t) \big[Q(t)\big]^{n-m}\;, \\ 
\beta_n(t) &= \sum\limits_{m=0}^n \binom{n}{m} Z_m(t) \big[-Q(t)\big]^{n-m}\;.
\end{split}
\label{eq:beta_Z_trans}
\end{equation}
For example, the first three order parameters are expressed with the newly introduced variables as
\begin{equation*}
\begin{aligned}
Z_1 &=Q+\beta_1\;,\qquad Z_2 = Q^2+2Q\beta_1+\beta_2\;,\\
& \quad Z_3 = Q^3+3Q^2\beta_1+3Q\beta_2+\beta_3\;.
\end{aligned}
\end{equation*}

\section{Reduction of the dynamics to three complex variables}
\label{sec:tcv}

As outlined in the previous section~\ref{sec:fdr}, there are many finite-dimensional invariant manifolds (with a finite
number of additional variables $\beta_n$) beyond the OA two-dimensional manifold (which corresponds to   vanishing $\beta_n$ for $n\geq 1$).
However, as already mentioned in Ref.~\onlinecite{cestnik_pikovsky_2022}, it is not excluded that different $\beta_n$ could be dependent.
Below we show that this is indeed the case, and the dynamics of the whole (even infinite) hierarchy
of variables $\beta_n(t)$ can be reduced to two complex equations.
\subsection{Six-dimensional reduction}
We now introduce two new complex variables $y(t),s(t)$ and new dynamical equations:
\begin{subequations}
\begin{align}
\dot{Q} &= (\ii\omega -\gamma) Q + h - h^* Q^2\;, \label{eq:eqQ_noise}\\
\dot{y} &= (\ii\omega - \gamma  - 2h^*Q) y\;, \label{eq:eqp_noise}\\
\dot{s} &= h^*y\;. \label{eq:eqs_noise}
\end{align}
\label{eq:zy_eqs}
\end{subequations}
Our goal below is to demonstrate, that these equations are equivalent to the infinite system~\eqref{eq:Qbeta} and therefore to the original system~\eqref{eq:Z_dyn}. At this point we would like to mention, that Eqs.~\eqref{eq:zy_eqs}
look like a skew system: variable $Q$ appears on the r.h.s. of \eqref{eq:eqp_noise}, and variable
$y$ appears on the r.h.s. of \eqref{eq:eqs_noise}, but variables $y,s$ do not appear on the r.h.s. of \eqref{eq:eqQ_noise}.
However, in most applications one describes a population with global coupling, where $\omega,h$ depend on the
order parameters $Z_n$, and thus on all dynamical variables $Q,y,s$ (cf. Eq.~\eqref{eq:moments_with_mu} below).

To show how this system represents dynamics~\eqref{eq:Qbeta}, we first introduce additional auxiliary variables $\alpha_n(t)$ by transforming $\beta_n(t)$: 
\begin{equation}
\beta_n(t) = y^n (t) \alpha_n (t)\;.
\label{eq:relation_br}
\end{equation}
We take the time derivative of this relation and divide both sides by $n \beta_n$
\[
\frac{\frac{1}{n}\dot{\beta}_n}{\beta_n} = \frac{\dot{y}}{y} + \frac{\frac{1}{n}\dot{\alpha}_n}{\alpha_n} \;,
\]
and then insert the dynamics of $\beta_n$~\eqref{eq:beta} and $y$~\eqref{eq:eqp_noise} :
\[
(\ii\omega-\gamma-2h^*Q) - h^*\frac{\beta_{n+1}}{\beta_n} = (\ii\omega-\gamma-2h^*Q) + \frac{\frac{1}{n}\dot{\alpha}_n}{\alpha_n} \;.
\]
Notice how the majority of the terms cancel, including all the effects of frequency $\omega(t)$ and noise $\gamma$.
As a result, the dynamics of the variables $\alpha_n(t)$ simplifies to:
\begin{equation}
\frac{1}{n} \dot{\alpha}_n = - h^* y\; \alpha_{n+1}\;.
\label{eq:Bn_dynamics}
\end{equation}

Now let us introduce the OGF of the variables $\alpha_n(t)$: 
\begin{equation}
\mathcal{A}(k,t) = \sum\limits_{n=1}^\infty \alpha_n(t) k^n\;,
\end{equation}
and express the dynamics \eqref{eq:Bn_dynamics} in terms 
of this OGF (see Appendix \ref{ap:gf} for the derivation):
\begin{equation}
\dot{\mathcal{A}} = -h^* y \Big[\mathcal{A}'-\frac{1}{k}\mathcal{A}\Big]\;.
\label{eq:R_dyn}
\end{equation}

Next we introduce yet another set of variables $\mu_n$.
This time we launch them not directly, but via an expression of the corresponding OGF $\mathcal{M}(k,t) = \sum\limits_{n=1}^\infty \mu_n k^n$ in terms of $\mathcal{A}(k,t)$:
\begin{equation}
\frac{\mathcal{M}(k)}{k} = \frac{\mathcal{A}(k+s)}{k+s}
\label{eq:transform}
\end{equation}
(where $s$ is the variable in \eqref{eq:eqs_noise}). 
By taking the time derivative of Eq.~\eqref{eq:transform}, we obtain:
\[
\frac{\dot{\mathcal{M}}(k)}{k} = \frac{\dot{\mathcal{A}}(k+s)}{k+s} + \dot{s} \Big[ \frac{\mathcal{A}'(k+s)}{k+s} - \frac{\mathcal{A}(k+s)}{(k+s)^2} \Big]\;.
\]
Now we insert the dynamics of $s$ according to Eq.~\eqref{eq:eqs_noise}, as well as the dynamics of $\mathcal{A}$ according to Eq.~\eqref{eq:R_dyn} 
and behold, the right-hand side of this relation is zero. This means that the OGF $\mathcal{M}$ and the corresponding 
variables $\mu_n$ are constant in time: $\dot{\mu}_n = 0$, $\dot{\mathcal{M}}(k) = 0$. In other words, the variables $\mu_n$
are integrals of motion.
This completes the proof that Equations~\eqref{eq:zy_eqs} are equivalent to Equations~\eqref{eq:Qbeta}. 

The radius of
convergence of the constant function $\mathcal{M}(k)$ is determined by the
asymptotic behavior of the coefficients $\mu_n$. Below we show that
these coefficients initially coincide with the order parameters $Z_n(0)$. Since they are bounded $|Z_n|\leq 1$, the radius of convergence of $\mathcal{M}$ is at least one. During dynamical evolution, the relevant argument of function $\mathcal{M}$ is $s(t)$, and in all our simulations we never observed it getting larger than one in absolute value: $|s(t)| < 1$. 

It is instructive to rephrase the relation~\eqref{eq:transform}, formulated above in terms of OGFs, to  the level of the variables 
(where it corresponds to a modified binomial transform, see Appendix \ref{sec:tranf_der} for the derivation):
\begin{equation}
\begin{split}
\mu_n &= \sum\limits_{m=n}^\infty \binom{m-1}{n-1} \alpha_m(t) \big[s(t)\big]^{m-n}\;, \\
\alpha_n(t) &= \sum\limits_{m=n}^\infty \binom{m-1}{n-1} \mu_m \big[-s(t)\big]^{m-n}\;.
\end{split}
\label{eq:BM_transform}
\end{equation}
Notice that we do not write the time argument of $\mu_n$ because these quantities are constants.

Using this relation, as well as how the variables $\alpha_n(t)$ relate to $\beta_n(t)$~\eqref{eq:relation_br}, and then how $\beta_n(t)$ relate to the order parameters $Z_n(t)$~\eqref{eq:beta_Z_trans}, we can express the order parameters in terms of the constant function $\mathcal{M}(k)$ and the three dynamic variables $Q(t),y(t),s(t)$ (see Appendix \ref{sec:moments_with_M} for the derivation) for all times (here we omit the time dependence in notation for convenience):
\begin{equation}
Z_n = Q^n - \sum\limits_{m=1}^n \binom{n}{m} Q^{n-m} y^m \sum\limits_{d=0}^{m-1} \frac{s^{d-m}}{d!} \mathcal{M}^{(d)} (-s)\;,
\label{eq:moments_with_mu}
\end{equation}
where $\mathcal{M}^{(d)}$ denotes the $d^\text{th}$ derivative of $\mathcal{M}$ with respect to $k$. 
In particular, the first circular moment (the Kuramoto order parameter) is expressed as:
\begin{equation}
Z_1 = Q-y \frac{\mathcal{M}(-s)}{s} \;.
\label{eq:first_moment}
\end{equation}
Notice how at $s=0$ one has to take the limit $\lim\limits_{\e\to0} \frac{\mathcal{M}(-\e)}{\e} = -\mu_1$. 
For all moment expressions~\eqref{eq:moments_with_mu} we have to consider 
similar limits: $\lim\limits_{\e\to0}\sum\limits_{d=0}^{m-1} \frac{\e^{d-m}}{d!}\mathcal{M}^{(d)}(-\e) = -\mu_m$. 
When performing numerical integration, one thus requires an expansion 
of the above expression~\eqref{eq:moments_with_mu} for small $Q$ and small $s$:
\begin{equation}
Z_n = y^n \left[\mu_n + n \mu_{n-1} \frac{Q}{y} - n  \mu_{n+1} s \right] + 
\mathcal{O}(Q^2, Qs, s^2) \;,
\label{eq:Z_expansion}
\end{equation}
we remind that $\mu_0 \equiv 1$ for normalization reasons. 
The need for expansion~\eqref{eq:Z_expansion} can also be avoided by considering a different definition of the constant function~\eqref{eq:transform}, see Appendix~\ref{sec:alternative_constant} - it also simplifies the expression for moments~\eqref{eq:moments_with_mu}. 

\subsection{Initial conditions}
\label{sec:ic}
The new set of variables $Q,y,s,\mu_n$ is not uniquely determined by the initial order parameters $Z_n$,
and in this section we discuss possible variants of determining them.
Different choices for the initial conditions of  $Q,y,s,\mu_n$ define the constant function $\mathcal{M}(k)$ differently. 
We first illustrate this with the simplest example of the OA manifold.

\subsubsection{Different choices of variables for OA initial conditions} 
As discussed above, on the OA manifold the EGF reads $\mathsf{Z}(k,t)=\exp[k Z(t)]$ and the order parameter $Z$
obeys $\dot Z=(\ii\omega -\gamma) Z+h-h^* Z^2$. Suppose, having a set of initial moments $Z_n(0)=Z^n(0)$, we
want to introduce new variables $Q,\beta_n$ according to \eqref{eq:zb}. One can immediately see
that a set $\beta_n(0)=\beta^n(0)$ is admissible if $Q(0)+\beta(0)=Z(0)$, in this case $\mathsf{B}(k,0)=\exp[k \beta(0)]$.
Furthermore, relation \eqref{eq:relation_br} allows for different choices of $y(0)$ and $\alpha_n(0)$. For 
any choice of $y(0)$,
we obtain $\alpha_n(0)=\alpha^n$, with $\alpha=\beta(0)/y(0)$. It is easy to see that independently of the choice
of $\beta(0)$ and $y(0)$, the dynamics of the order parameters is the same. Indeed, in this case $\mathcal{M}(k)=\alpha k/(1-\alpha k)$
and the main order parameter according to \eqref{eq:first_moment} is $Z_1=Q+\alpha y/(1+\alpha s)$. Calculation of the 
derivative $\dot Z_1$ from the general dynamical equations \eqref{eq:zy_eqs} yields the correct equation  $\dot Z_1=(\ii\omega -\gamma) Z_1+h-h^* Z_1^2$, i.e. the system remains on the OA manifold. In this specific case of pure OA dynamics, it is natural to consider $\alpha = 0$ such that $\mathcal{M} = 0$ and the only relevant equation is the OA equation \eqref{eq:eqQ_noise}~\cite{pikovsky_rosenblum_2008,pikovsky_rosenblum_2011}. 

\subsubsection{Variant A: A simple choice of initial variables}
\label{sec:icsimp}
Here we present possibly the simplest choice of initial conditions for $Q,y,s,\mu_n$. We initially set $Q$ and $s$ to zero and set $y$ to 1, so that all the variable 
sets $\mu_n, \alpha_n, \beta_n$ and $Z_n$ coincide:
\begin{equation}
\begin{split}
&Q(0) = 0\;,\\
&y(0) = 1\;,\\
&s(0) = 0\;,\\
&\mu_n = \alpha_n(0) = \beta_n(0) = Z_n(0)\;.
\end{split}
\label{eq:init_cond}
\end{equation}
Then the constant function can simply be determined by the initial order parameters: 
\begin{equation}
\mathcal{M}(k) = \sum\limits_{n=1}^\infty Z_n(t=0)\, k^n\;. 
\label{eq:M_A}
\end{equation}
It is instructive to express this function in terms of the initial distribution of the phases
\begin{equation}
\mathcal{M}(k) = \int_0^{2\pi} d\vp\;P(\vp,0)\frac{ke^{\ii\vp}}{1-ke^{\ii\vp}}\;. 
\label{eq:M_Ad}
\end{equation}

As discussed above, this is not the only possible choice of initial conditions and in some cases might not be optimal.
Notice that for this choice, the OA manifold corresponds to the   
constants being  powers of $Z_1(0)$: $\mu_n = Z^n_1(0)$, so they do not vanish, 
as is commonly considered~\cite{pikovsky_rosenblum_2008,pikovsky_rosenblum_2011}. In terms of dynamics, the two descriptions are equivalent.

Next we list specific functions $\mathcal{M}(k)$ for some examples of the initial states in
this variant A of the initial conditions~\eqref{eq:init_cond}. We stress here, that a representation 
via an initial distribution density is valid for the interpretation of the system with identical
oscillators under Cauchy white noise (section \ref{sec:pfcn}). For the case of non-identical oscillators with a distribution
of frequencies (section \ref{sec:pfcf}), one should operate with the order parameters directly.
\begin{itemize}
\item A uniform distribution $P(\vp,0) = \frac{1}{2\pi}$ corresponds to $\mathcal{M}(k) = 0$. Here all the moments vanish, 
which is a trivial invariant state of the dynamics~\eqref{eq:continuity}. 
\item A delta distribution of the phases $P(\vp,0)=\delta(\vp-\vp_0)$ corresponds to $Z_n=\exp(\ii n\vp_0)$ and thus $\mathcal{M}(k) = \frac{e^{\ii \varphi_0}\,k}{1-e^{\ii \varphi_0}\,k}$.  
\item A wrapped Cauchy distribution $P(\vp,0) = \frac{1}{2\pi} \frac{1-|\mu|^2}{|1-\mu e^{-\ii\vp}|^2}$
with a complex parameter $\mu \in \mathbb{C}$, 
corresponds to $\mathcal{M}(k) = \frac{\mu k}{1-\mu k}$. The moments are  powers of the parameter $\mu$: 
$\mu_n = \mu^n$, which means this state is on the OA manifold~\cite{ott_antonsen_2008}. 
\item A Kato-Jones distribution~\cite{kato-jones} $P(\vp,0) = \frac{1}{2\pi} (1+c\frac{\mu e^{-\ii\vp}}{1-\mu e^{-\ii\vp}} + \text{c.c.})$ with complex parameters $c,\mu \in \mathbb{C}$, corresponds to $\mathcal{M}(k) = c\frac{\mu k}{1-\mu k}$. 
Its moments are described by a power series multiplied with a complex constant: $\mu_n = c\, \mu^n$, making it a skewed/asymmetric generalization of the wrapped Cauchy distribution.
\item Distributions with a finite number of moments 
$P(\vp,0) = \frac{1}{2\pi} (1+\sum\limits_{n=1}^N \mu_n e^{-\ii n\vp}+\text{c.c.})$
 correspond to polynomials $\mathcal{M}(k) = \sum\limits_{n=1}^N \mu_n k^n$. 
\item Distributions with binomial moments $\mu_n = \binom{n}{m} \mu^n$ for $m \geq 1$ correspond to rational functions $\mathcal{M}(k) = \frac{(\mu k)^m}{(1-\mu k )^{m+1}}$. Their phase density reads: $P(\vp,0) = \frac{1}{2\pi}(1+\frac{\mu^m e^{-\ii m \vp}}{(1-\mu e^{-\ii\vp})^{m+1}}+\text{c.c.})$. 
\item A half-uniform distribution 
\begin{equation}
P(\vp,0) = \begin{cases} \frac{1}{\pi} & \text{if}\ \vp \in (\vp_0,\vp_0+\pi)\;,\\0 & \text{else}\;,\end{cases}
\label{eq:half_uni}
\end{equation}
corresponds to $\mathcal{M}(k) = \frac{2\ii}{\pi}\arctanh(e^{\ii\vp_0}k)$. Here odd moments are fractions: $\mu_{2n-1} = \frac{2\ii}{\pi} \frac{\exp(\ii(2n-1)\vp_0)}{2n-1}$, 
and even ones are equal to zero: $\mu_{2n} = 0$, $n\geq1$. 
\item A sawtooth distribution $P(\vp,0) = \frac{1}{\pi}(1-\frac{\vp-\vp_0}{\pi}+ \lfloor \frac{\vp-\vp_0}{\pi}\rfloor) $ 
corresponds to $\mathcal{M}(k) = -\frac{\ii}{\pi}\log(1-e^{\ii 2\vp_0}k^2)$. Here even moments are fractions: $ \mu_{2n} = \frac{2\ii}{\pi}\frac{\exp(\ii2n\vp_0)}{2n}$, 
and odd ones are equal to zero: $\mu_{2n-1} = 0$, $n\geq1$. 
\end{itemize}

In practice, the phase density can only be observed empirically, and therefore it may not be clear how to choose $\mathcal{M}(k)$. In such situations it can always be approximated with a finite truncation of its Taylor series: $\mathcal{M}(k)\approx \sum\limits_{n=1}^N Z_n(0) k^n$, one only needs to estimate the first few initial moments $Z_n(0)$. This just corresponds to approximating the initial state with a finite Fourier representation. In Appendix~\ref{sec:M_finite} we present a numerical example where we test the convergence of such approximations. 

We end this subsection with the following remark: if the initial distribution of the phases is a weighted sum
of ``elementary'' distributions $P(\varphi,0)=\sum_m c_m P_m(\varphi,0)$ with real weights $c_m \in \mathbb{R}$ that add up to 1: $\sum_m c_m=1$ (additionally
one should ensure that $P(\varphi,0)\geq 0$),  then the constant generating function is the 
weighted sum of the corresponding 
``elementary'' generating functions $\mathcal{M}(k)=\sum_m c_m \mathcal{M}_m(k)$. In particular, 
in Refs.~\onlinecite{engelbrecht_mirollo_2020,Ichiki-Okumura-20} 
a superposition of several wrapped Cauchy distributions has been considered as an initial state;
in terms of the approach above this corresponds to $\mathcal{M}(k)=\sum_m c_m \frac{\varkappa_m k}{1-\varkappa_m k}$,
where complex parameters $\varkappa_m$ characterize partial distributions.

\subsubsection{Variant B: Initial conditions based on the OA manifold}
\label{sec:icoa}
Often  initial states that are close to the OA manifold are of interest. Suppose that the order parameters are 
well described as powers of a complex constant, with minor perturbations:
\begin{equation}
Z_n(0) = R^n + \e_n\;. 
\label{eq:perturbation}
\end{equation} 
In this case a different initial condition appears natural:
\begin{equation}
\begin{split}
&Q(0) = R\;,\\
&y(0) = 1\;,\\
&s(0) = 0\;,\\
&\mu_n = \alpha_n(0) = \beta_n(0) = \sum\limits_{m=1}^n \binom{n}{m}\e_m (-R)^{n-m}\;,
\end{split}
\label{eq:init_cond_pert}
\end{equation}
and the constant function $\mathcal{M}(k)$ is expressed as:
\begin{equation}
\mathcal{M}(k) = \sum\limits_{n=1}^\infty k^n \sum\limits_{m=1}^n \binom{n}{m} \e_m (-R)^{n-m}\;.
\label{eq:M_B}
\end{equation}
This function is small if values of $\e_n$ are small. Notice however, that this ``perturbation'' approach is actually global, 
because smallness of $\e_n$ is not needed, and $\mathcal{M}(k)$ need not be small for this description to be valid. 
Note that the definition of function $\mathcal{M}(k)$ depends on the choice of initial conditions (cf. \eqref{eq:M_A} and \eqref{eq:M_B}), so the list of specific $\mathcal{M}(k)$ functions in Section~\ref{sec:icsimp} does not apply here. The moments are still described with Eq.~\eqref{eq:moments_with_mu} but for numerical integration one needs to expand them beyond~\eqref{eq:Z_expansion} for only small $s$ (such expansions can be avoided by considering an alternative constant function to $\mathcal{M}$, see Appendix~\ref{sec:alternative_constant}). 

A simple specific example where the perturbation to the OA manifold
in \eqref{eq:perturbation}, is one where only the first harmonic term is perturbed: $\e_n=0$ for $n>1$. In this case $\mathcal{M}(k) = \e_1 \frac{k}{(1+Rk)^2}$ and so the first moment is given by $Z_1 = Q + \e_1 \frac{y}{(1-Rs)^2}$. The dynamics follow~\eqref{eq:zy_eqs}.

We mention here that in some cases, an extension of the set of variables might be appropriate. As an example,
we show in Appendix \ref{sec:ap} a possibility to describe an initial
state similar to~\eqref{eq:perturbation} with a system of four complex variables.

\subsection{Finite ensemble numerical comparison}

\begin{figure}[!htb]
\centering
\includegraphics[width=0.99\columnwidth]{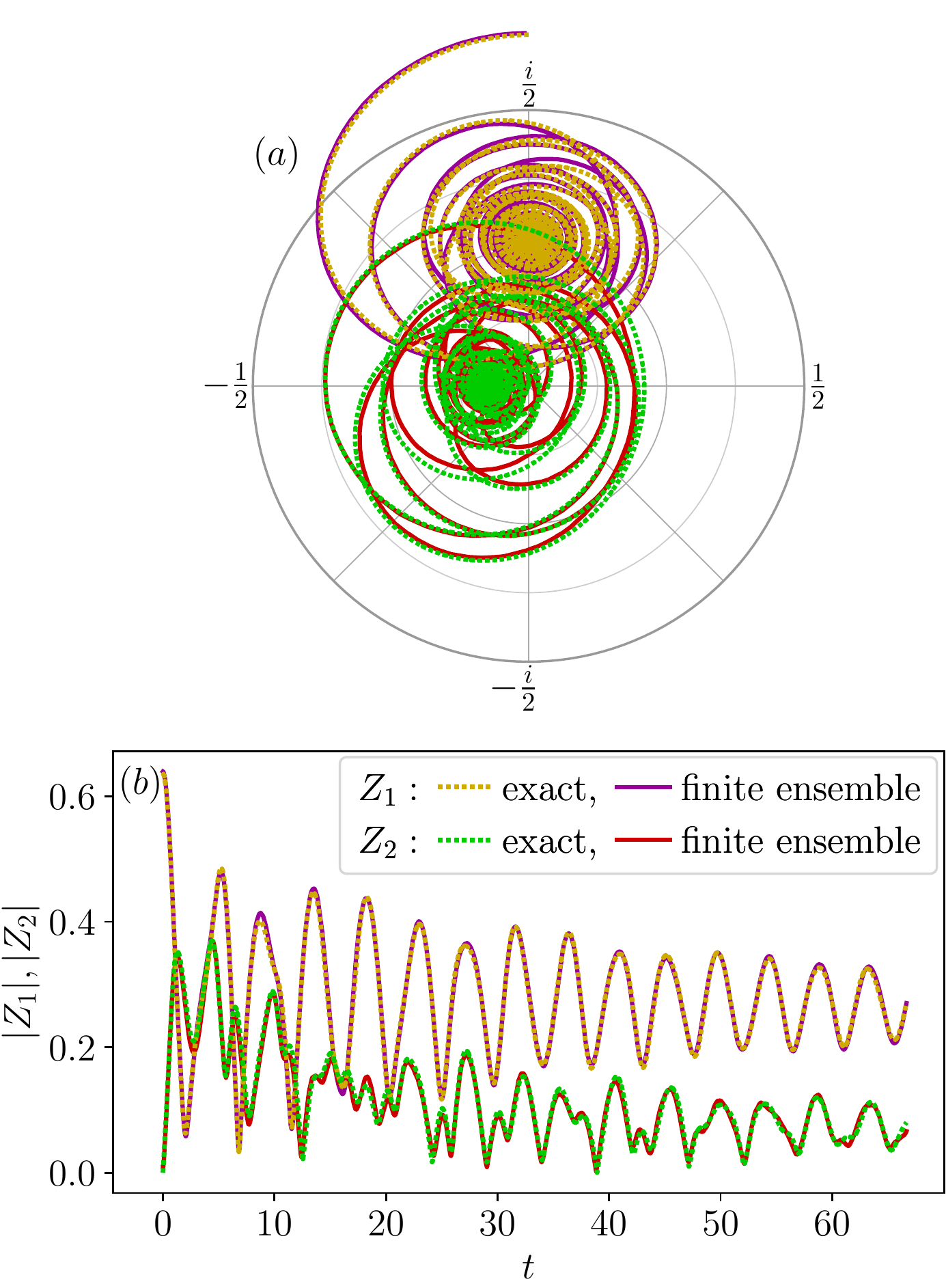}
\caption{Comparison of the finite ensemble dynamics of $10\,000$ Josephson phases~\eqref{eq:jj1} with the low-dimensional dynamics in the thermodynamic limit~\eqref{eq:zy_eqs}.   In panel $(a)$ the first two circular moments are compared in the complex plane, and in panel $(b)$ their absolute values are compared as functions of time. Solid lines: simulation of a finite ensemble; dotted lines (which practically overlap with solid lines): solution of exact equations in the thermodynamic limit. }
\label{fig:finite_comparison}
\end{figure}

Here we numerically compare the derived low-dimensional dynamics~\eqref{eq:zy_eqs}, which is exact in the thermodynamic limit, with a simulation of a finite ensemble. We take the example already explored in Ref.~\onlinecite{cestnik_pikovsky_2022}: an array of overdamped noisy Josephson junctions coupled via a resistive load~\cite{Watanabe-Strogatz-94}. The equations for the Josephson phases read
\begin{equation}
\dot\varphi_j=1 + a\sin(\varphi_j)+\frac{\e}{N}\sum_{n=1}^N\sin(\varphi_n)+ \gamma \xi_j(t)\;.
\label{eq:jj1}
\end{equation}
In terms of the basic model \eqref{eq:phase_system_noise}, this corresponds to the choice of $\omega,h$:
$\omega=1+\e\,\text{Im}[Z_1]$ and $h=-\frac{a}{2}$. 
We consider parameters: $a=-0.7$, $\e=1.5$, and noise strength $\gamma = 0.02$. 
To highlight the advantage of the new derivation, we consider the initial distribution of phases to be half-uniform~\eqref{eq:half_uni}, thus starting far from the OA manifold, and also far from initial states that are easy to represent in terms of the $Q,\beta_n$ hierarchy~\cite{cestnik_pikovsky_2022}. 
For the finite ensemble we consider $10\,000$ phases, which are initially randomly sampled in the interval $[0,\pi)$. For our exact reduction in the thermodynamic limit,
we take initial conditions from variant A~\eqref{eq:init_cond}: $Q(0) = s(0) = 1-y(0) = 0$, and therefore the constant function $\mathcal{M}(k) = \frac{2\ii}{\pi}\arctanh(k)$. The comparison of the trajectories  in Fig.~\ref{fig:finite_comparison} shows a  very close match, and we expect it to be even closer for larger ensembles. The discrepancies appear to be only due to finite size effects, we stress that in the limit of an infinitely large ensemble our reduction~\eqref{eq:zy_eqs} is exact.

\section{Global stability of the OA manifold}
\label{sec:gsoa}

Stability of the OA manifold has been discussed in Refs.~\onlinecite{Ott-Antonsen-09,pietras_daffertshofer_2016,engelbrecht_mirollo_2020}.
Here we demonstrate how these results are reproduced in our approach. To show the attractiveness of the OA manifold it
is enough to demonstrate that the variable $y$ tends to zero $y\to 0$. Indeed, for $y=0$ we have from \eqref{eq:relation_br}
$\beta_n=0$, $n\geq 1$, and from \eqref{eq:zb} it follows that the solution is on the OA manifold.

Let us introduce two new variables $Y,S$ according to relations
\begin{equation}
y=(1-|Q|^2)Y\;,\qquad s=Q^* S\;.
\label{eq:YS}
\end{equation}
The equations for these variables read 
\begin{subequations}
\begin{align}
\dot Y&=\left(\ii\omega+hQ^*-h^*Q-\gamma\frac{1+|Q|^2}{1-|Q|^2}\right)Y\;,
\label{eq:Y}\\
\dot S&=\Big(\ii\omega +hQ^*-h^*Q+\gamma\Big)S+\left(h^*Q-\frac{h^*}{Q^*}
\right)(S-Y)\;.\label{eq:S}
\end{align}
\label{eq:YSeq}
\end{subequations}
Here we focus on the equation for $Y$, and will use Eq.~\eqref{eq:S} in Section~\ref{sec:nfc} below.
From \eqref{eq:Y} we obtain the following evolution of $|Y|$:
\begin{equation}
\frac{d|Y|}{dt} \frac{1}{|Y|} =-\gamma\frac{1+|Q|^2}{1-|Q|^2}\leq -\gamma\;.
\label{eq:modY}
\end{equation}
The latter inequality on the r.h.s. follows from the property $0\leq |Q|^2<1$. Indeed, the equation for
the evolution of $|Q|^2$ reads
\[
\frac{d}{dt}|Q|^2=-2\gamma|Q|^2+(hQ^*+h^*Q)(1-|Q|^2)
\]
and at $|Q|=1$ the derivative $\frac{d}{dt}|Q|^2=-2\gamma$ is negative for $\gamma>0$,
thus this boundary is not reachable. 

Integrating inequality \eqref{eq:modY} yields
\begin{equation}
|Y(t)|\leq |Y(0)|\,e^{-\gamma t}\;,
\label{eq:Y_bound}
\end{equation}
and consequently, since $|y| \leq |Y|$:
\begin{equation}
|y(t)|\leq \frac{|y(0)|}{1-|Q(0)|^2} \,e^{-\gamma t}\;,
\label{eq:y_bound}
\end{equation}
which means that $|Y|$ and $|y|$ vanish exponentially fast in time. 
This proves the attractiveness of the OA manifold for system \eqref{eq:Z_dyn} for $\gamma>0$.

\begin{figure}[!htb]
\centering
\includegraphics[width=0.99\columnwidth]{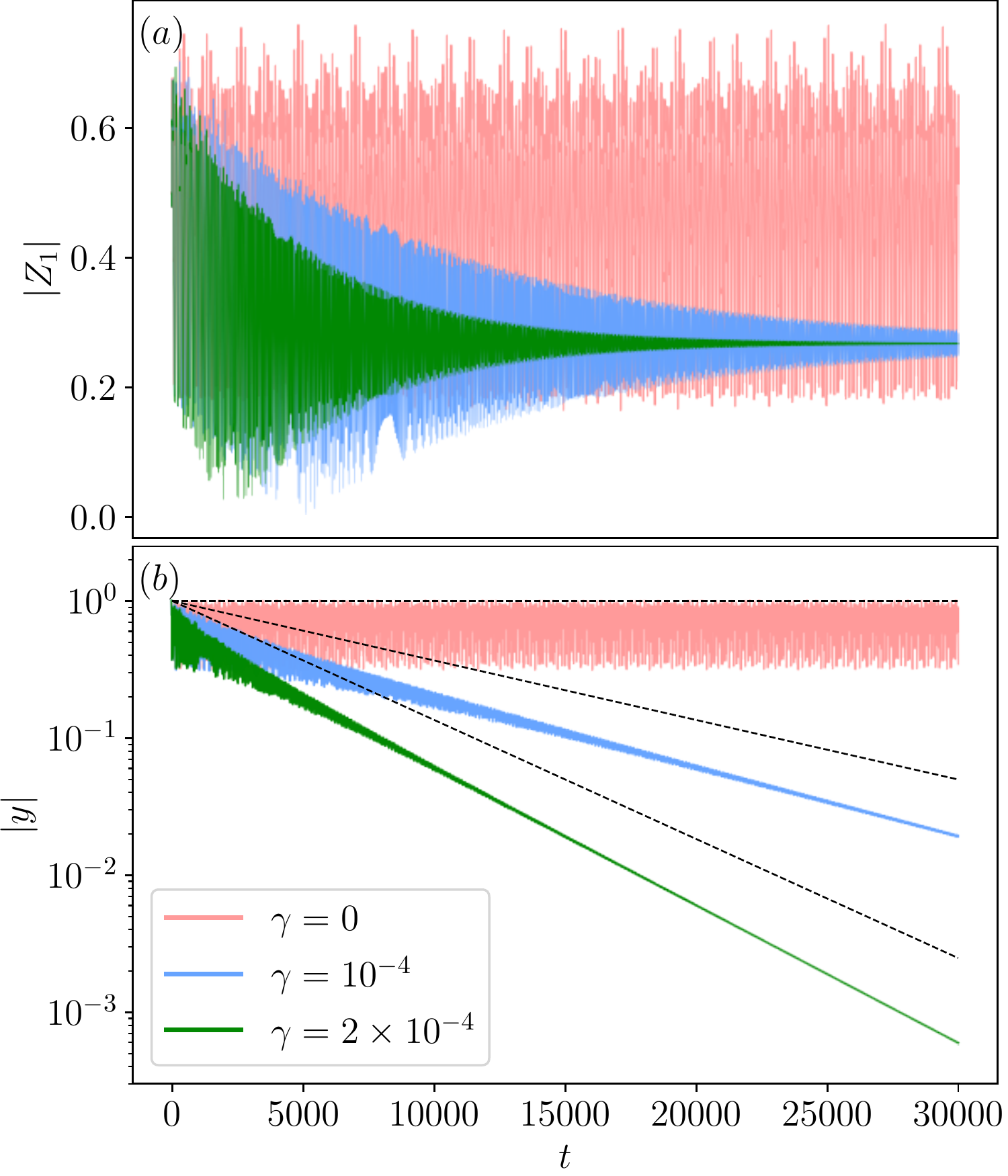}
\caption{Time series of the coupled Josephson junctions array~\eqref{eq:jj1} for three different noise strengths; $\gamma = 0$ depicted with red, $\gamma = 10^{-4}$ depicted with blue and $\gamma = 2\times10^{-4}$ depicted with green. Panel $(a)$: the norm of the first order parameter $Z_1$; panel $(b)$: the norm of variable $y$. The initial conditions for all three cases are the same: $Q(0) = s(0) = 1-y(0) = 0$ (variant A) and constant function $\mathcal{M}(k) = 0.4 k$, which means the initial distribution has only one harmonic. In the case of no noise, chaotic dynamics is observed, while in the noisy case there is an initial chaotic stage which is followed by a clear exponential decay of the $y$ variable. In panel $(b)$ the upper bound~\eqref{eq:y_bound} for all three cases is shown with a dashed black line. }
\label{fig:jj}
\end{figure}

\subsection{Example}
Here we illustrate stability
of the OA manifold numerically. 
We take the already explored example of Josephson junctions~\eqref{eq:jj1}, with the same 
parameters $a=-0.7$, $\e=1.5$. 
In Ref.~\onlinecite{cestnik_pikovsky_2022} it was demonstrated that for $\gamma=0$, the dynamics outside of the OA manifold is chaotic. 
In Fig.~\ref{fig:jj} we again demonstrate chaotic behavior 
in this system for $\gamma=0$, and a transition 
to regular dynamics for $\gamma=10^{-4}$ and $\gamma=2\times10^{-4}$. %Runge-Kutta $4^\text{th}$-order method with timestep $10^{-4}$ was used for integrating dynamics~\eqref{eq:zy_eqs}. 
The exponential decay of $|y|$, which is bounded by~\eqref{eq:y_bound} is evident at large times.

\section{Noise-free case and a relation to the Watanabe-Strogatz theory}
\label{sec:nfc}
Watanabe and Strogatz~\cite{watanabe_strogatz_1993,Watanabe-Strogatz-94} demonstrated that a population of identical noiseless oscillators can be 
reduced to three real dynamical variables plus constants of motion. To see that this case is included
in our theory,
let us consider identical oscillators and no noise, thus taking $\gamma = 0$. 

It is instructive to start with Eqs.~\eqref{eq:YSeq}. It is easy to see, that
for $\gamma=0$, the manifold $S=Y$ is invariant:
\[
\frac{d}{dt}(S-Y)=\left[\ii\omega+hQ^*-\frac{h^*}{Q^*}\right](S-Y)\;.
\]
Using the definition of variables $Y, S$~\eqref{eq:YS}, the manifold is described in terms of $y$ and $s$ as:
\begin{equation}
Y = \frac{y}{1-|Q|^2} = \frac{s}{Q^*} = S\;.
\end{equation}
Since we initially set $Q(0)=s(0)=1-y(0)=0$ for arbitrary states, one can always
set initial conditions on this manifold such that $Y(0)=S(0)=1$. Thus, two equations \eqref{eq:YSeq} reduce to one.
Moreover, because of \eqref{eq:modY}, for $\gamma=0$  the variable $Y$ remains on the unit circle for all times,
and we can introduce an angle variable $\theta(t)=\text{arg}(Y(t))$. 
Its evolution follows from \eqref{eq:YSeq}:
\begin{equation}
\dot{\theta} = \omega  -\ii( hQ^* - h^*Q)\;.
\label{eq:theta_dyn}
\end{equation}
This angle variable $\theta$ just corresponds to the WS angle variable, while $Q$ is the WS 
order parameter~\cite{watanabe_strogatz_1993,Watanabe-Strogatz-94}. The two equations~\eqref{eq:eqQ_noise} 
(with $\gamma=0$) and \eqref{eq:theta_dyn} then represent the exact evolution. We explicitly derive the equivalence
with the WS approach in Appendix \ref{sec:wsr}.

In the WS theory, the relation between the original phases $\varphi_j(t)$ and constant phases $\psi_j$ is given by the 
M\"obius transform \cite{Watanabe-Strogatz-94,Marvel-Mirollo-Strogatz-09}:
\begin{equation}
e^{\ii \vp_j} = \frac{e^{\ii (\psi_j+\theta)}+Q}{1+Q^* e^{\ii (\psi_j+\theta)}} \; , \qquad e^{\ii (\psi_j+\theta)} = \frac{e^{\ii \vp_j}-Q}{1-Q^* e^{\ii \vp_j}} \;.
\label{eq:mobius}
\end{equation}
Our choice of the initial conditions $Q(0)=\theta(0)=0$ (it corresponds to the ``identity conversion'' in terms of WS,
cf. Eq. (5.10) in Ref.~\onlinecite{Watanabe-Strogatz-94}) means that $\psi_j=\vp_j(0)$. Thus, because
the integrals $\mu_n$ are defined as $\mu_n=Z_n(0)$, these quantities are the circular moments
of the transformed constant phase variables in the WS approach  $\mu_n = \langle e^{\ii n\psi} \rangle$.

Watanabe and Strogatz have shown that these transformations are  also valid for a finite number of oscillators,
but this case is not covered by our approach.
We mention here, that in the WS formalism there is also a freedom in choosing the order parameter $Q$ and the phase variable $\theta$;
this freedom is similar to the one discussed in Section \ref{sec:ic} above.

\section{Lyapunov spectrum}
\label{sec:ls}

Our theory describes the dynamics outside of the OA manifold, and is thus suitable for consideration
of small perturbations transversal to this manifold. Such perturbations define the Lyapunov spectrum of the dynamics, together
with the perturbations tangential to this manifold. The system of equations \eqref{eq:Qbeta} is most suitable for this
analysis. The OA manifold corresponds to vanishing $\beta_n$, therefore Eqs.~\eqref{eq:beta} define the transversal perturbations.
Since these equations are a skew system, each $\beta_n$ defines two Lyapunov exponents (because $\beta_n$ are complex).
One can straightforwardly derive from \eqref{eq:beta}, omitting the skew term $\sim \beta_{n+1}$ on the r.h.s.,
the averaged evolution for the magnitude of a perturbation:
\[
\frac{1}{2n}\av{\frac{d}{dt} \ln |\beta_n|^2}=\av{-\gamma-h^*Q-hQ^*}=\Lambda\;.
\] 
Thus, the Lyapunov spectrum consists of the exponents within the OA manifold (which are calculated
using linearised Eq.~\eqref{eq:Q}), and of doubly degenerated values $n\Lambda$, $n=1,2,3,\ldots$.

\section{Response of the Ott-Antonsen regime to a resetting}
\label{sec:res}

As has been already discussed in the literature~\cite{Ott-Antonsen-09,pietras_daffertshofer_2016,engelbrecht_mirollo_2020} and in Section \ref{sec:gsoa}, in the system of equations \eqref{eq:zy_eqs}
the OA manifold is attracting if $\gamma > 0$ (at least in the weak sense, but because we follow only
the moments of the phase distribution, such an attraction is enough). 
In terms of variables $Q,y,s$ with a nontrivial constant
function $\mathcal{M}(k)$, this corresponds to $y\to 0$ as $t\to\infty$ (see Section \ref{sec:gsoa}). 
For the conservative case $\gamma = 0$, see Section \ref{sec:nfc} above.  

The approach above allows for calculating the evolution from an arbitrary state to the OA manifold via solutions of \eqref{eq:zy_eqs}. One can reformulate such a problem as a resetting one: One starts with the dynamics on the OA manifold; then an instant ``resetting'' to a state outside of this manifold is performed.
The evolution of \eqref{eq:zy_eqs} then shows what will be the final state after re-attraction to the OA manifold.  A particular question of interest 
depends on the type of the attractors on the OA manifold. If there is only one global attractor, then the trajectory returns to it. If this 
attractor is periodic or quasiperiodic, the returning trajectory will be phase shifted with  respect to the unperturbed one (in the 
quasiperiodic case one expects phase shifts in every direction of independent oscillations). Here one speaks about a phase resetting or a 
phase response curve (PRC)~\cite{Canavier-06,smeal2010phase}. For a chaotic global attractor, generally one does not expect a resetting to have a drastic effect (although for strange 
attractors with well-defined phase variables a phase resetting similar to the periodic case can be defined~\cite{Schwabedal-Pikovsky-Kralemann-Rosenblum-12}; it can lead to phase 
synchronization of chaos if periodically repeated \cite{Pikovsky-Rosenblum-Osipov-Kurths-97}). In the case of multistability, the 
most drastic effect of resetting would be a jump to another basin of attraction, so that the final state will be another attractor on the OA 
manifold (in case of multistable periodic attractors one can additionally follow the phase 
response~\cite{Grines-Osipov-Pikovsky-18}). Below we consider several examples, for small and large resettings.

\subsection{Perturbation theory in terms of an (infinitesimal) PRC}
Suppose we have a state on the OA manifold with a complex order parameter $R$, so that $\langle e^{\ii n\vp}\rangle=R^n$.
Let us apply to all the phases a transformation
\begin{equation}
\vp\to\vp+\e f(\vp)\;,
\label{eq:phtr}
\end{equation}
where $f(\vp)=\sum_m f_m e^{\ii m\vp}$ is a PRC function (given by its Fourier representation) and $\e \ll 1$ is assumed to be small. Let us calculate the circular moments just after a resetting, in order $\e$:
\begin{gather*}
Z_n=\av{e^{\ii n(\vp+\e f(\vp))}}\approx \av{e^{\ii n \vp}(1+\ii n\e f(\vp))}= \\
=R^n+\ii n \e  \sum\limits_{m=-\infty}^\infty f_m \av{e^{\ii (n+m)\vp}}\;.
\end{gather*}
Since $m$ can be negative, calculation of the latter average is not a simple expression, because
\[
\av{e^{\ii (n+m)\vp}}=\begin{cases} R^{n+m} &n+m\geq 0\;,\\
(R^*)^{|n+m|} & n+m<0\;,
\end{cases}
\]
therefore we restrict ourselves to two simplest cases.
\subsubsection{First harmonics resetting}
In this case $f(\vp)=f_1e^{\ii \vp}+f_1^* e^{-\ii\vp}$.
For $n\geq 1$ we have $n+m\geq 0$ and therefore for both $m=\pm 1$
we can write $
\av{e^{\ii (n+m)\vp}}=R^{n+m}
$.
Thus
\[
Z_n=R^n(1+\ii n\e (f_1R+f_1^*R^{-1}))\;,\qquad n\geq 0\;.
\]
Calculation of the EGF yields
\[
\mathsf{Z}(k,0)=e^{kR}(1+\ii\e k(f_1R^2+f_1^*))\;,
\]
where we used $\sum_n n \frac{x^n}{n!}=xe^{x}$.

Let us now transform to variables $Q,y,s$ and take $Q(0)=R$ (like in variant B, Section~\ref{sec:icoa}).
This means that the EGF $\mathsf{B}$ is
\[
\mathsf{B}(k,0)=1+\ii\e k(f_1R^2+f_1^*)\;.
\]
We come to the conclusion, that only one variable $\beta_1$ is non-zero, and the system can be directly and exactly solved with variables $Q,\beta_1$ by virtue of Eqs.~\eqref{eq:Qbeta}; there is no need to go to the full system \eqref{eq:zy_eqs}. Alternatively, one can consider only the first two equations of system~\eqref{eq:zy_eqs} 
and function $\mathcal{M}(k) = \beta_1(0) k$ and then the third variable $s$ does not matter. If we rewrite Eqs.~\eqref{eq:Qbeta} in terms of variables $(Z_1,\beta_1)$, we obtain
\begin{equation*}
\begin{aligned}
\dot Z_1&=(\ii\omega -\gamma)Z_1+h-h^*Z_1^2+h^*\beta_1^2\;,\\
\dot \beta_1&=(\ii\omega-\gamma-2h^*(Z_1-\beta_1))\beta_1\;.
\end{aligned}
\label{eq:Zbeta}
\end{equation*}
One can see that the correction to the standard OA equation is $h^*\beta_1^2\sim \e^2$. Thus,
in the first order in $\e$, inclusion of the additional variable $\beta_1$ is irrelevant and the resetting is well described within the OA equation.

\subsubsection{Second harmonics resetting}
In this case $f(\vp)=f_2e^{\ii 2\vp}+f_2^* e^{-\ii2\vp}$, and we have
$\sum_m f_m \av{e^{\ii (n+m)\vp}}=f_2\av{e^{\ii(n+2)\vp}}+
f_2^* \av{e^{\ii (n-2)\vp}}$. Thus, the term with $n=1$ reads $f_2R^3+f_2^* R^*$,
while all the higher-order terms $n\geq 2$ can be written in a unified way $f_2 R^{n+2}+f_2^* R^{n-2}$.
Rewriting the term with $n=1$ as
$
f_2R^3+f_2^*R^{-1}+[f_2^*R^*-f_2^*R^{-1}]
$, we
 obtain
\[
Z_n=R^n+\ii \e n R^n[f_2R^2+f_2^* R^{-2}]+\ii\e \delta_{n,1}[f_2^*R^*-f_2^*R^{-1}]\;,
\]
where $\delta_{n,1}$ is the Kronecker delta.
This yields the following EGF
\[
\mathsf{Z}(k,0)=
e^{kR}(1+\ii \e k(f_2 R^3+f_2^* R^{-1}))+ \ii\e k[f_2^*R^*-f_2^*R^{-1}]\;.
\]
Now the EGF $\mathsf{B}(k,0)$ is nontrivial, choosing $Q(0)=R$:
\[
\mathsf{B}(k,0)=
1+\ii \e k (f_2 R^3+f_2^* R^{-1})+\ii\e ke^{-kR} [f_2^*R^*-f_2^*R^{-1}]\;.
\]
This allows for obtaining a closed expression of the 
constant function (for choice $y(0)=1,\;s(0)=0$) as 
\[
\mathcal{M}(k)=\ii\e k\left[f_2R^3+f_2^*\frac{R^*+2k+ R k^2}{(1+Rk)^2}\right]\;.
\]
After this, system~\eqref{eq:zy_eqs} is to be solved.

\subsection{Large resettings}
Unfortunately, a transformation of the type \eqref{eq:phtr} is hardly tractable for large $\e$. Here we
discuss another way of resetting, which leads to closed expressions even for large changes of the phases. This approach is applicable to identical oscillators subject to Cauchy white noise, but not
for the distribution of natural frequencies.

Suppose, in the OA state with order parameters $Z_n=R^n$,
we randomly  choose a portion $\e$ of all oscillators and reset 
them completely (this means that they ``forget'' their old states), cf. Ref.~\onlinecite{Gupta-resetting}. We consider two variants below.

\subsubsection{Random resetting}
Here we assume that the new phases in the affected set become uniformly distributed in the interval $[0,2\pi)$.
These oscillators do not contribute to new order parameters which thus take the values $Z_n=(1-\e)R^n$.
This corresponds to an offset Cauchy distribution, or specifically, a weighted superposition of the Cauchy distribution and the uniform distribution. 
If one uses variant A of initial conditions, then evolution
starts from the initial values $Q(0)=s(0)=0$, $y(0)=1$ and $\mathcal{M}(k)=(1-\e)Rk/(1-Rk)$ is determined via Eq.~\eqref{eq:M_A}. Alternatively,
adopting variant B, one can start from the initial conditions $Q(0)=R$, $y(0)=1$, $s(0)=0$ and then $\mathcal{M}(k)=-\e Rk/(1+Rk)$ is determined via Eq.~\eqref{eq:M_B}. Then the system evolves according to Eqs.~\eqref{eq:zy_eqs}. 

However, due to the simplicity of this example, there is an even easier way of treating this situation. Notice how 
moments can be viewed as a superposition of two OA contributions, referred to as Poisson kernels by Ref~\onlinecite{engelbrecht_mirollo_2020}: 
\begin{equation}
Z_n = (1-\e)Q_1 + \e Q_2\;,
\label{eq:two_kernels}
\end{equation}
where initially $Q_1(0) = R$ and $Q_2(0) = 0$. 
In this case therefore, one can evolve the system by considering two OA equations~\eqref{eq:Q}, which only interact through the forcing $h$, and the solution maintains the form~\eqref{eq:two_kernels} for all times.  

\subsubsection{Coherent resetting} 
\label{sec:res-cr}
Consider now that reset phases are not distributed uniformly but rather take on another distribution $P^\text{(res)}(\vp)$. If this distribution is a wrapped Cauchy (which includes the uniform and the delta distribution) the setting can again be treated simply as a superposition of the OA modes (a.k.a. Poisson kernels)~\footnote{Resetting with a Kato-Jones~\cite{kato-jones} distribution (asymmetric generalization of the wrapped Cauchy) can be accomplished with two additional OA modes~\cite{cestnik_pikovsky_2022}.}. However, the reset distribution can generally have a different form. Below we consider two cases.

Case (i): Partially coherent resetting.
Here the reset phases are distributed according to a single harmonic 
density: $P^\text{(res)}(\vp) = \frac{1}{2\pi} \left[1 + 2c\cos(\vp-\vp_0)\right]$, $c,\vp_0\in\mathbb{R}$. The reset distribution therefore has only one non-zero moment: $Z_1^\text{(res)} = ce^{\ii\vp_0}$, and the full distribution after a portion $\e$ of phases are reset is described by: $Z_n = (1-\e) R^n + \e c e^{\ii\vp_0} \delta_{n,1}$. 
If using variant A initial conditions, the variables initialize as $Q(0) = s(0) = 0$, $y(0) = 1$ and $\mathcal{M}(k) = (1-\e)\frac{Rk}{1-Rk}+\e c e^{\ii\vp_0}k$ according to Eq.~\eqref{eq:M_A}. Alternatively if choosing variant B, then $Q(0) = R$, $y(0) = 1$, $s(0) = 0$ and $\mathcal{M}(k) = -\e\left[ \frac{Rk}{1+Rk}+ \frac{ce^{\ii\vp_0} k}{(1+Rk)^2}\right]$ according to Eq.~\eqref{eq:M_B}. Then the system evolves following Eqs.~\eqref{eq:zy_eqs}. 
One could also treat this setting as one OA mode and one general contribution described by full Eqs.~\eqref{eq:zy_eqs}, we provide 
such a description in Appendix~\ref{sec:ap}.

Case (ii): Fully coherent resetting. Here  the reset phases take the same value $\vp_0$. The new order parameters are thus
$Z_n=(1-\e)R^n+\e e^{\ii n \vp_0}$. Again both variants of initializing the variables after the reset are possible. In variant A one sets $Q(0)=s(0)=0$, $y(0)=1$ and $\mathcal{M}(k)=(1-\e)\frac{Rk}{1-Rk}+\e
\frac{e^{\ii \vp_0}k}{1- e^{\ii \vp_0}k}$ is determined via Eq.~\eqref{eq:M_A}, while following variant B, we can set
$Q(0)=R$, $y(0)=1$, $s(0)=0$ and $\mathcal{M}(k) = \e  \frac{(e^{\ii \vp_0}-R)k}{1-(e^{\ii \vp_0}-R)k}$ is determined via Eq.~\eqref{eq:M_B}. 
As mentioned before, since the delta distribution is a special case of the wrapped Cauchy, this case could also be described by just two OA modes~\eqref{eq:two_kernels}. 

The expressions above can be readily extended to a setup where several randomly chosen subpopulations
of oscillators are reset with different distributions. Such an approach has been discussed in the context of application
to de-synchronization of neurons for Parkinson patients~\cite{Tass-99}. 

\begin{figure}[!htb]
\centering
\includegraphics[width=0.99\columnwidth]{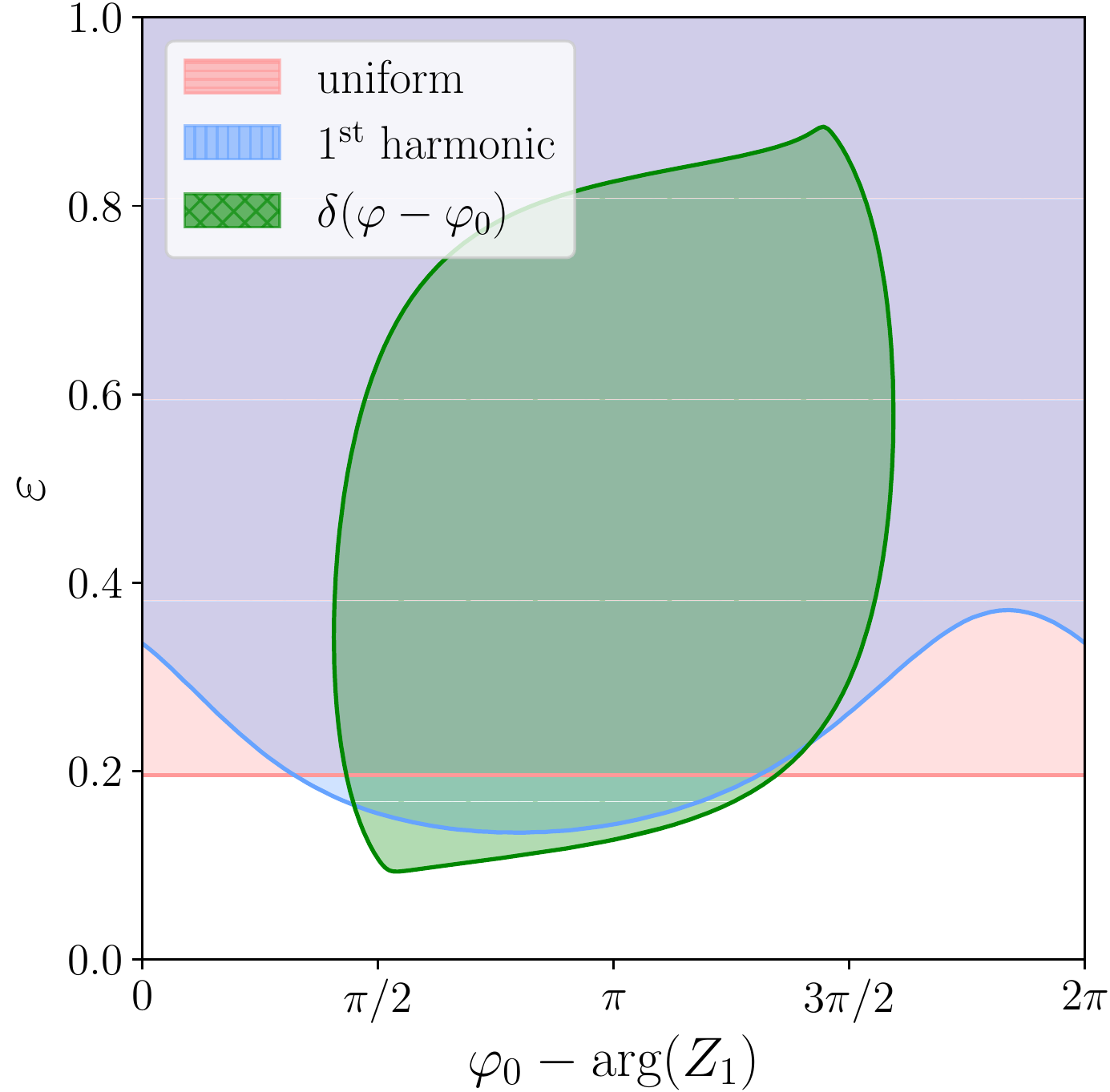}
\caption{Switching domains for large resetting in the bistable system~\eqref{eq:bistable}. Both the synchronous and asynchronous regimes are stable. Starting from the synchronous regime on the OA manifold, we reset an $\e$ portion of phases according to three example distributions: uniform (red domain), case (i) -- single harmonic density with amplitude $c=0.5$  (blue domain), and case (ii) -- delta distribution (green domain). The shaded regions mark areas that induce a switch to asynchrony. }
\label{fig:cr}
\end{figure}

\subsubsection{Numerical example for large resettings}

As an example we consider a simple Kuramoto-type system with a synchrony-asynchrony 
bistability~\cite{Pikovsky-Rosenblum-09}. In this setup $\omega=\text{const}$ (and one can without loss of generality set
this parameter to zero), and force is 
\begin{equation}
h=Z_1 \exp[\ii\theta_0+\ii\theta_1 |Z_1|^2]\;.
\label{eq:bistable}
\end{equation}
We set  $\gamma=0.1$, $\theta_0=0.8\pi$ 
and $\theta_1=4$. For these parameters the states with $Z_1=0$ and $|Z_1|\approx 0.948$ are both stable. 
We start with the latter state of a nearly synchronized ensemble, and apply the three types of resetting as described above.  By solving the reduced six-dimensional equations~\eqref{eq:zy_eqs}, we obtain the domain of parameters $\e,\vp_0$ for which the resettings lead to a transition to the asynchronous state $Z_1=0$. 

For the random resetting, there is no dependence on $\vp_0$, and the corresponding domain is $\e>0.196$ (above red line in Fig.~\ref{fig:cr}). For the coherent resetting, we consider two cases discussed above, (i) and (ii), and 
their corresponding basins are depicted in Fig.~\ref{fig:cr} with blue and green domains, respectively. One can see that in all 
three cases, a finite
perturbation is needed to suppress synchrony. For the coherent resettings, there is an optimal combination 
of $\e$ and $\vp_0$; the coherent subpopulation should be phase shifted around $\pi$ relative to the phase
of the mean field of the non-reset units. For case (ii) we also see that if $\e$ is too large, the reset units form a new cluster and the synchrony remains.

\section{Conclusion}
\label{sec:concl}

First, we summarize the approach and findings of this paper. Our starting point is an infinite
system of equations for the circular moments (order parameters). These equations contain damping due to
either Cauchy white noise, or a Cauchy distribution of natural frequencies. 
By virtue of several transformations, which 
are formulated in terms of generating functions, we reduce this system to three complex equations. Additionally, a complex-valued function of one variable is defined, which remains constant during the evolution. 
The order parameters at each moment of time are represented through this function
and the three complex dynamical variables. 

The original set of equations for the order parameters have the same form in two situations: 
if the phase oscillators are subject to a Cauchy white noise, and if the natural frequencies
are Cauchy distributed (but time-independent). Only in the former case there is a simple unique correspondence between the order parameters and the distribution of the phases. In the latter situation,
one can calculate the order parameters from the distribution of the phases (under the assumption of analyticity of the density in the upper complex plane of frequencies), but it appears impossible to
reconstruct this distribution from the order parameters without further assumptions. Therefore, the results of the paper are fully
applicable to noisy ensembles, but some approaches (e.g., phase resetting) are not suitable for the oscillators with distributed frequencies.

The theory includes both the WS description (noise-free identical oscillators) and the OA manifold (on which 
the dynamical variable $y$  vanishes). In the framework of our approach,
one can simply demonstrate that  the dynamical variable $y(t)$ tends to zero, 
which corresponds to the weak stability of the OA manifold discussed in the literature. Therefore, our approach
is an essential improvement compared to OA theory, if a transient evolution from an initial state
outside of the OA manifold is important. In particular, it allows for a calculation of the full basins 
of different attractors lying on the OA manifold.

In this paper we operated with the phase equations. In some cases it is convenient to transform
the phase equations to other variables (e.g., theta-neurons, equations which belong to class
\eqref{eq:phase_system_noise}~\cite{Luke-Barreto-So-13,Laing-14}, can be transformed to so-called quadratic integrate-and-fire neurons
\cite{laing_2015,montbrio_pazo_roxin_2015,bick_goodfellow_laing_martens_2020}). An extension of the theory
to quadratic integrate-and-fire neurons will be presented elsewhere~\cite{Pietras-Cestnik-Pikovsky-22}.

\acknowledgments
We thank L. Smirnov and R. Toenjes for useful discussions. The work was supported by DFG (grant No. PI 220/21-1).

\appendix

\section{From the dynamics of the moments to the PDEs for the generating functions}
\label{ap:gf}

Consider a sequence of variables $\{f_n\}$ 
(the only condition is that the generating functions below do exist; because we apply the theory
to bounded circular moments, this appears to always be the case). 
Also, in our case $f_0\equiv1$.

We distinguish between two types of generating functions: exponential generating function (EGF) defined as
\begin{equation}
\mathsf{F}(k) = \sum\limits_{n=0}^\infty f_n \frac{k^n}{n!}\;,
\end{equation}
and ordinary generating function (OGF) defined as
\begin{equation}
\mathcal{F}(k) = \sum\limits_{n=1}^\infty f_n k^n\;.
\end{equation}
Here we show how the dynamics of the variables $f_n$, given by an infinite set of ODEs,
 can be translated to the dynamics in terms of generating functions $\mathsf{F}$ and $\mathcal{F}$, 
 given by a single partial differential equation (PDE). 
Suppose the dynamics of $f_n$ is as follows:
\begin{equation}
\dot{f}_n =n\left( a\,f_{n-1}+b\, f_{n} + c\, f_{n+1}\right)\;, \quad n \geq 1\;,
\label{eq:dyn}
\end{equation}
where $a,b,c$ are arbitrary complex quantities. Then the PDE for the EGF $\mathsf{F}$ reads:
\begin{equation}
\dot{\mathsf{F}} = a\,k\mathsf{F}+b\, k\mathsf{F}' + c\, k\mathsf{F}''\;.
\label{eq:egf_dyn}
\end{equation}
In fact, in the case of an EGF $\mathsf{F}(k,t)$, any term $f_{n+m}$ for $m \geq -1$ on the right-hand side of~\eqref{eq:dyn} just corresponds to a term of the form $k\mathsf{F}^{(m+1)}$ in equation~\eqref{eq:egf_dyn}.

To prove this, it is sufficient to express derivatives of $\mathsf{F}$ as formal series:
\begin{equation}
\begin{aligned}
\dot{\mathsf{F}} &= \sum\limits_{n=0}^\infty \dot{f}_n \frac{k^n}{n!}\;,\\
\mathsf{F} &= \sum\limits_{n=0}^\infty f_n \frac{k^n}{n!} = \frac{1}{k} \sum\limits_{n=0}^\infty (n+1)\, f_n \frac{k^{n+1}}{(n+1)!} = \\ 
&= \frac{1}{k} \sum\limits_{n=-1}^\infty (n+1)\, f_n \frac{k^{n+1}}{(n+1)!} = \frac{1}{k} \sum\limits_{n=0}^\infty n\, f_{n-1} \frac{k^{n}}{n!}\;,\\
\mathsf{F}' &= \sum\limits_{n=1}^\infty f_n \frac{k^{n-1}}{(n-1)!} = \frac{1}{k} \sum\limits_{n=1}^\infty n\, f_n \frac{k^{n}}{n!} = \frac{1}{k} \sum\limits_{n=0}^\infty n\, f_n \frac{k^{n}}{n!}\;,\\
\mathsf{F}'' &= \sum\limits_{n=2}^\infty f_n \frac{k^{n-2}}{(n-2)!} = \frac{1}{k} \sum\limits_{n=2}^\infty (n-1)\, f_n \frac{k^{n-1}}{(n-1)!} = \\
&= \frac{1}{k} \sum\limits_{n=1}^\infty n\, f_{n+1} \frac{k^{n}}{n!} = \frac{1}{k} \sum\limits_{n=0}^\infty n\, f_{n+1} \frac{k^{n}}{n!}\;,\\
\mathsf{F}^{(m)} &= \sum\limits_{n=m}^\infty f_n \frac{k^{n-m}}{(n-m)!} = \\
&= \frac{1}{k}\sum\limits_{n=m}^\infty (n-m+1)\, f_n \frac{k^{n-m+1}}{(n-m+1)!} = \\
&= \frac{1}{k}\sum\limits_{n=1}^\infty n\, f_{n+m-1} \frac{k^{n}}{n!} = \frac{1}{k}\sum\limits_{n=0}^\infty n\, f_{n+m-1} \frac{k^{n}}{n!}\;.
\end{aligned}
\label{eq:F_expansions}
\end{equation}
Notice that the sum index changes in the derivation, e.g. if we have the summed terms proportional to $n$ and the sum starts with index ``1'': $\sum\limits_{n=1} n\, (...)$, we can just add the zero term (since it vanishes) and write the sum from index ``0''. The way we have derived the above expressions, the right-most forms of~\eqref{eq:F_expansions} all correspond to a sum starting at $n=0$ and have a factor of $\frac{k^n}{n!}$, therefore we can just compare the terms under the sum to see that 
the dynamics \eqref{eq:dyn} corresponds to the PDE \eqref{eq:egf_dyn}.

More generally, a
term in the dynamics of the variables $f_n$ having the form
\begin{equation}
\dot{f}_n = ... + c\, n\, f_{n+m} + ... \;, \quad m \geq -1\;,
\end{equation}
in terms of the EGF $\mathsf{F}(k,t)$ corresponds to the partial derivative in the equation 
for $\mathsf{F}$:
\begin{equation}
\dot{\mathsf{F}} = ... + c\, k \mathsf{F}^{(m+1)} + ...
\end{equation}

Now let us consider the OGF $\mathcal{F}(k,t)$. 
This generating function is applicable to systems with only the homogeneous term $f_n$ and 
one higher term $f_{n+1}$ on the r.h.s., i.e. to the case $a=0$ in \eqref{eq:dyn}.
Let us write the relevant sums:
\begin{equation}
\begin{aligned}
\dot{\mathcal{F}} &= \sum\limits_{n=1}^\infty \dot{f}_n k^{n} \;,\\
k\mathcal{F}' &= \sum\limits_{n=1}^\infty n\, f_n k^n\;,\\
\mathcal{F}'-\frac{1}{k}\mathcal{F} &= \sum\limits_{n=1}^\infty n\, f_n k^{n-1} - f_n k^{n-1} = \\
&= \sum\limits_{n=1}^\infty (n-1)\, f_n k^{n-1} = \sum\limits_{n=1}^\infty n\, f_{n+1} k^n\;.
\end{aligned}
\label{eq:F_expansions2}
\end{equation}
Again, by comparing the right-most expressions of~\eqref{eq:F_expansions2}, we can see that the dynamics
\begin{equation}
\dot{f}_n = b\, n\, f_{n} + c\, n\, f_{n+1}  \;,
\end{equation}
in terms of the OGF $\mathcal{F}$, corresponds to
\begin{equation}
\dot{\mathcal{F}} = b\, k \mathcal{F}' + c\, \Big[\mathcal{F}'-\frac{1}{k} \mathcal{F} \Big]\;.
\end{equation}

\section{Transformation (\ref{eq:transform}) in terms of variables}
\label{sec:tranf_der}
We rewrite here relation \eqref{eq:transform}:
\begin{equation}
\frac{\mathcal{M}(k)}{k} = \frac{\mathcal{A}(k+s)}{k+s}\;.
\end{equation}
Expressed as an OGF series (with a different index for the right-hand side), it reads:
\begin{equation*}
\sum\limits_{n=1}^\infty \mu_n k^{n-1} = \sum\limits_{r=1}^\infty \alpha_r (k+s)^{r-1}\;.
\end{equation*}
Next, we expand the binomial term on the r.h.s.:
\begin{equation*}
\sum\limits_{n=1}^\infty \mu_n k^{n-1} = \sum\limits_{r=1}^\infty \alpha_r \sum\limits_{d=0}^{r-1} \binom{r-1}{d} k^{r-d-1} s^d\;,
\end{equation*}
gather together terms for which $r-d-1 = n-1$, and use the same index $n$ as on the l.h.s.:
\begin{equation*}
\sum\limits_{n=1}^\infty \mu_n k^{n-1} = \sum\limits_{n=1}^\infty k^{n-1} \sum\limits_{m=n}^\infty \binom{m-1}{m-n} \alpha_m s^{m-n}\;.
\end{equation*}
Now it is clear that this transformation corresponds to \eqref{eq:BM_transform}:
\begin{equation}
\mu_n = \sum\limits_{m=n}^\infty \binom{m-1}{m-n} \alpha_m s^{m-n}\;,
\end{equation}
where the binomial coefficient $\binom{m-1}{m-n}$ can also be written as $\binom{m-1}{n-1}$.

\section{Expressing moments $Z_n$ with the constant function $\mathcal{M}(k)$}
\label{sec:moments_with_M}

The $d^\text{th}$ derivative in $k$ of $\mathcal{M}(k)$ is expressed as:
\begin{equation}
\mathcal{M}^{(d)}(k) = \frac{d!}{k^d} \sum\limits_{m=1}^\infty \binom{m}{d} \mu_m k^m\;.
\end{equation}
Next, using the binomial relation: $\binom{m-1}{n-1} = (-1)^{n-1} \sum\limits_{d=0}^{n-1} (-1)^d \binom{m}{d}$ and the relation between $\alpha_n$ and $\mu_n$ variables~\eqref{eq:BM_transform}, we can write:
\begin{equation}
\alpha_n = - \sum\limits_{d=0}^{n-1} \frac{s^{d-n}}{d!} \mathcal{M}^{(d)}(-s)\;.
\end{equation}
Then we can use relation~\eqref{eq:relation_br} to express $\beta_n$ as:
\begin{equation}
\beta_n = -y^n \sum\limits_{d=0}^{n-1}\frac{s^{d-n}}{d!} \mathcal{M}^{(d)}(-s)\;, \quad n \geq 1 \;,
\end{equation}
and we can use \eqref{eq:beta_Z_trans} to express the moments:
\begin{equation}
Z_n = Q^n - \sum\limits_{m=1}^n \binom{n}{m} Q^{n-m} y^m \sum\limits_{d=0}^{m-1} \frac{s^{d-m}}{d!} \mathcal{M}^{(d)} (-s)\;.
\label{eq:moments_app}
\end{equation}

For initial condition variant A: $Q(0) = s(0) = 0$, $y(0) = 1$, to see that expression~\eqref{eq:moments_app} is correct initially at $t = 0$, we have to evaluate the following limits:
\begin{equation}
\lim\limits_{\e\to0}\sum\limits_{d=0}^{m-1} \frac{\e^{d-m}}{d!}\mathcal{M}^{(d)}(-\e) = -\mu_m\;,
\end{equation}
which confirm that $Z_n(0) = \mu_n$.

\section{Alternative perturbation around the OA manifold}
\label{sec:ap}

There are different ways one can write states close to the OA manifold. An alternative to the main text~\eqref{eq:perturbation} is:
\begin{equation}
Z_n = (1-\e)Z^n + \e\,p_n\;, \quad |\e| \ll 1 \;,
\end{equation}
Notice that due to the linearity of Eq.~\eqref{eq:F}, we can view this simply as splitting the state into two parts: one on and one off the OA manifold: $Z_n^{(\text{OA})} = Z^n$ and $Z_n^{(\text{pert})} = p_n$. 
Each of the two parts can then be treated separately, the only thing connecting them is the force $h$. We know that the part on the OA manifold requires only one complex equation~\eqref{eq:eqQ_noise} by considering $\mathcal{M}^{(\text{OA})} = 0$, while the rest can be treated as before with the full set of three equations~\eqref{eq:zy_eqs} and the simplest general initial conditions~\eqref{eq:init_cond}. The complete dynamics is then written as:
\begin{equation}
\begin{split}
\dot{Z} &= (\ii\omega -\gamma) Z + h - h^* Z^2\;, \\
\dot{Q} &= (\ii\omega -\gamma) Q + h - h^* Q^2\;, \\
\dot{y} &= (\ii\omega - \gamma  - 2h^*Q) y\;, \\
\dot{s} &= h^*y\;, 
\end{split}
\label{eq:zy_eqs_pert_alt}
\end{equation}
where the perturbations $Z_n^{(\text{pert})} = p_n$ express with $Q,y,s$ via Eq.~\eqref{eq:moments_with_mu}. Global order parameters are simply the weighted sum of both contributions, e.g. the first order parameter is expressed as: $Z_1 = (1-\e)Z + \e\Big( Q -y\frac{\mathcal{M}(-s)}{s} \Big)$, where $\mathcal{M} = \sum\limits_{n=1}^\infty p_n(t=0) k^n$. Notice that this perturbation description is actually global, $\e$ does not need to be small.

\section{Alternative constant function $\mathcal{N}(k) = \frac{\mathcal{M}(-k)}{-k}$}
\label{sec:alternative_constant}

Instead of transformation~\eqref{eq:transform} we can consider the following:
\begin{equation}
\mathcal{N}(k) = \frac{\mathcal{A}(s-k)}{s-k} = \frac{\mathcal{M}(-k)}{-k}\;.
\label{eq:alt_transform}
\end{equation}
The OGF 
$\mathcal{N}(k)$ is also a constant function in time, its time derivative yields:
\[
\dot{\mathcal{N}}(k) = \frac{\dot{\mathcal{A}}(s-k)}{s-k} + \dot{s} \Big[ \frac{\mathcal{A}'(s-k)}{s-k} - \frac{\mathcal{A}(s-k)}{(s-k)^2} \Big] \ = 0\;.
\]
The benefit of this function is that it simplifies the expression for Kuramoto-Daido order parameters~\eqref{eq:moments_with_mu}:
\begin{equation}
Z_n = Q^n \Big[1 - \sum\limits_{m=1}^n \binom{n}{m} \Big(-\frac{y}{Q}\Big)^m \frac{\mathcal{N}^{(m-1)} (s)}{(m-1)!}\Big]\;,
\label{eq:moments_with_mu_alternative}
\end{equation}
the first three moments explicitly:
\begin{equation*}
\begin{aligned}
Z_1 &= Q+y \mathcal{N}(s) \;,\\
Z_2 &= Q^2 + 2 Q y \mathcal{N}(s) - y^2\mathcal{N}'(s)\;,\\
Z_3 &= Q^3 + 3 Q^2y\mathcal{N}(s) - 3Qy^2\mathcal{N}'(s) + y^3\frac{\mathcal{N}''(s)}{2}\;,
\end{aligned}
\end{equation*}
and no expansions for small $s$ like~\eqref{eq:Z_expansion} are needed. 

If we consider this function $\mathcal{N}(k)$ as the OGF of some variables $\nu_n$ we now have to also consider the {\it zero}$^\text{th}$ term: 
\begin{equation}
\mathcal{N}(k) = \sum_{n=0}^\infty \nu_n k^n\;.
\end{equation}
(notice $n$ starting from 0). 
Constant quantities $\nu_n$ express with $\mu_n$ as:
\begin{equation}
\nu_n = (-1)^n \mu_{n+1}\;.
\end{equation}
Considering variant A initial conditions~\eqref{eq:init_cond}: $Q(0) = s(0) = 0,\ y(0) = 1$, these quantities can be determined as:
\begin{equation}
\nu_n = (-1)^n Z_{n+1}(0)\;,
\end{equation}
and so the constant function $\mathcal{N}(k)$ can be expressed as:
\begin{equation}
\mathcal{N}(k) = \sum_{n=0}^\infty (-1)^n Z_{n+1}(0) k^n\;.
\end{equation}

\section{Approximating $\mathcal{M}(k)$ with a finite series}
\label{sec:M_finite}

\begin{figure}[!htb]
\centering
\includegraphics[width=0.9\columnwidth]{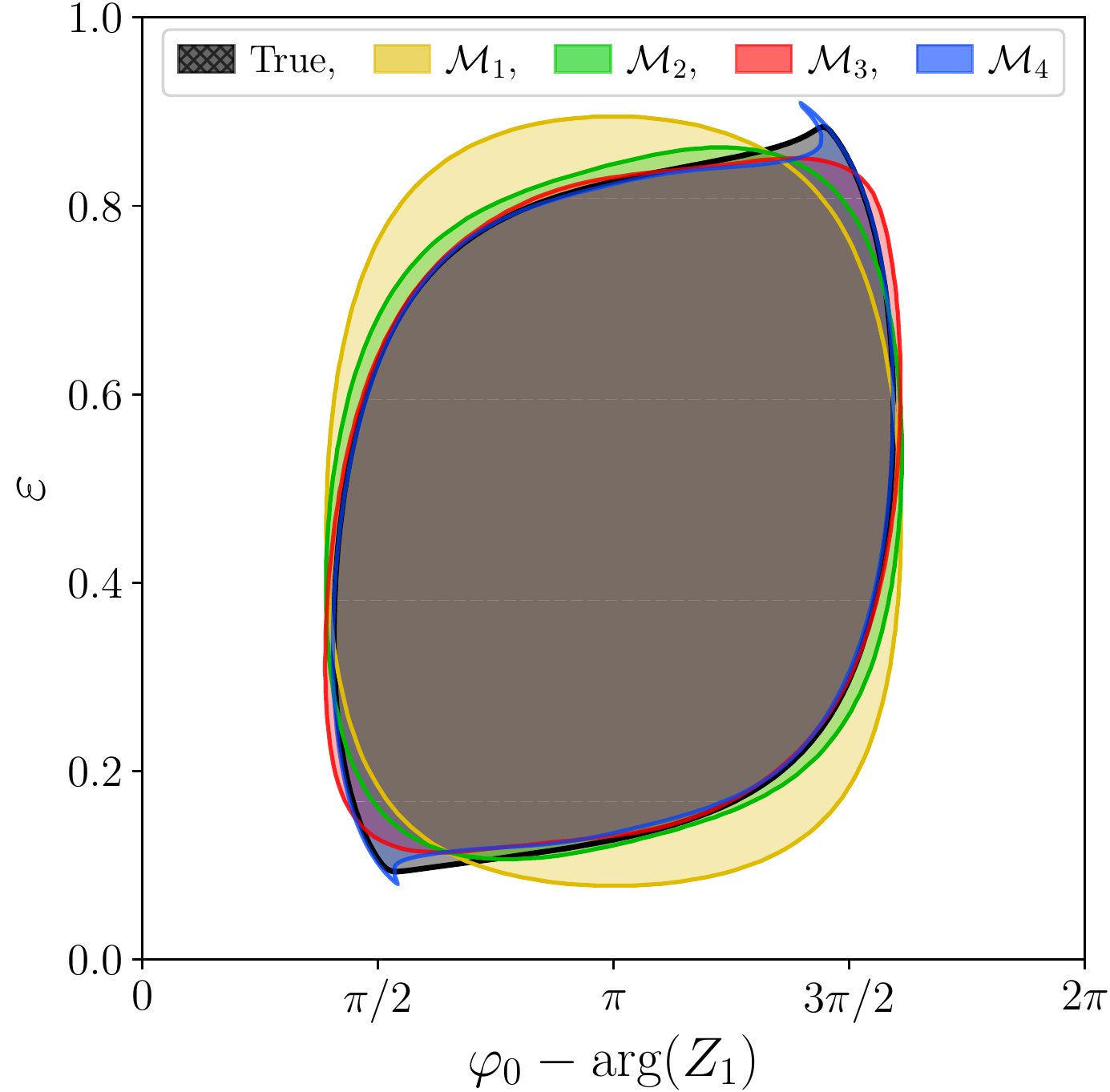}
\caption{The switching domains for resetting the bistable system~\eqref{eq:bistable} with a $\delta$-distribution, for different approximations the constant function $\mathcal{M}(k)$ (the black cross-shaded region corresponds to the green region in Fig.~\ref{fig:cr}). The considered approximations are truncations of its Taylor series at different numbers of terms: $\mathcal{M}_N = \sum\limits_{n=1}^N Z_n(0)k^n$. }
\label{fig:M_series}
\end{figure}

For empirically observed phase densities it might not always be clear how to determine the constant function $\mathcal{M}(k)$, but one can always estimate the first few moments $Z_n(0)$ numerically and then approximate the function with a series: $\mathcal{M}(k) \approx \sum\limits_{n=1}^N Z_n(0) k^n$. 
Here  in Fig.~\ref{fig:M_series} we show an example of that. 
We consider the bistable system~\eqref{eq:bistable} and just like in Fig.~\ref{fig:cr} compute the switching domain for resetting with a $\delta$-distribution. Several different truncations of the series are considered: $N = 1, 2, 3, 4$, and the corresponding domains are depicted in Fig.~\ref{fig:M_series}. We see a fast convergence of the approximations.

\section{Relation to the Watanabe-Strogatz theory}
\label{sec:wsr}
In this appendix we demonstrate explicitly that the derived equations coincide with the Watanabe-Strogatz equations
for identical oscillators. Our basic equation for the oscillator dynamics \eqref{eq:phase_system_noise} in the noise-free case
reads
\[
\dot \varphi=\omega+\text{Im}[2he^{-\ii\varphi}]=\omega+2\text{Im}[h]\cos\varphi-2\text{Re}[h]\sin\varphi\;.
\]
Watanabe and Strogatz~(Eq. (5.3) in Ref.~\onlinecite{Watanabe-Strogatz-94}) write this equation in slightly different notations
\[
\dot\varphi=F+G\cos\varphi+H\sin\varphi\;,
\]
so that $F=\omega$, $G=2\text{Im}[h]$, $H=-2\text{Re}[h]$. After a transformation (Eq. (5.4) in Ref.~\onlinecite{Watanabe-Strogatz-94})
\[
\tan\left(\frac{\varphi-\Phi}{2}\right)=\sqrt{\frac{1+\eta}{1-\eta}}\tan\left(\frac{\psi-\Psi}{2}\right)\;,
\]
they obtain a set of equations for the variables  $\eta,\Phi,\Psi$ (Eq. (5.9) in Ref.~\onlinecite{Watanabe-Strogatz-94})
\begin{equation}
\begin{aligned}
\dot\eta&=-(1-\eta^2)(G\sin\Phi-H\cos\Phi) \;,\\
\eta\dot\Phi&=\eta F-G\cos\Phi-H\sin\Phi \;,\\
\eta\dot\Psi&=-\sqrt{1-\eta^2}(G\cos\Phi+H\sin\Phi) \;.
\end{aligned}
\label{eq:ws1}
\end{equation}
The variables $\psi$ are constants of motion.

Let us introduce a new variable $\rho$ according to
\[
\frac{1-\rho}{1+\rho}=\sqrt{\frac{1+\eta}{1-\eta}}\;.
\]
Then
$\eta=-\frac{2\rho}{1+\rho^2}$, $1-\eta^2=\frac{(1-\rho^2)^2}{(1+\rho^2)^2}$,
$\sqrt{1-\eta^2}=\frac{1-\rho^2}{1+\rho^2}$, and $\dot\eta=-2\dot\rho\frac{1-\rho^2}{(1+\rho^2)^2}$.
We substitute these relations in \eqref{eq:ws1} and obtain
\begin{equation}
\begin{aligned}
\dot\rho&=\frac{1-\rho^2}{2}(G\sin\Phi-H\cos\Phi) \;,\\
\dot\Phi&=F+\frac{1+\rho^2}{2\rho}(G\cos\Phi+H\sin\Phi) \;,\\
\dot\Psi&=\frac{1-\rho^2}{2\rho}(G\cos\Phi+H\sin\Phi) \;.
\end{aligned}
\label{eq:ws2}
\end{equation}

Let us now introduce variables $Q=\rho e^{\ii\Phi}$ and $\theta=\Phi-\Psi$
and assume that these variables fulfill Eqs. \eqref{eq:eqQ_noise} 
(with $\gamma=0$), \eqref{eq:theta_dyn} above:
\begin{equation}
\begin{aligned}
\dot Q&=\ii\omega Q+h-h^*Q^2 \;,\\
\dot\theta&=\omega-\ii(hQ^*-h^*Q) \;.
\end{aligned}
\label{eq:cp1}
\end{equation}
The equations for $\rho,\Phi,\Psi$ following from \eqref{eq:cp1} read:
\begin{equation}
\begin{aligned}
\dot \rho&=(1-\rho^2)(\text{Re}[h]\cos\Phi+\text{Im}[h]\sin\Phi) \;,\\
\dot\Phi&=\omega+\frac{1+\rho^2}{\rho}(\text{Im}[h]\cos\Phi-\text{Re}[h]\sin\Phi) \;,\\
\dot\Psi&=\frac{1-\rho^2}{\rho}(\text{Im}[h]\cos\Phi-\text{Re}[h]\sin\Phi) \;.
\end{aligned}
\label{eq:cp2}
\end{equation}
Substituting here $\omega=F$, $\text{Re}[h]=-\frac{H}{2}$ and $\text{Im}[h]=\frac{G}{2}$, we obtain
exactly system \eqref{eq:ws2}, which proves the equivalence of our equations \eqref{eq:eqQ_noise} 
(with $\gamma=0$), \eqref{eq:theta_dyn} and the WS equations \eqref{eq:ws2}.

% ===================================

%

\end{document}